\documentclass[aps,preprint,nofootinbib]{revtex4}%
\usepackage{amsfonts}
\usepackage{amsmath}
\usepackage{amssymb}
\usepackage{graphicx}%
\providecommand{\U}[1]{\protect\rule{.1in}{.1in}}

\begin{document}
\title{Darboux coordinates for the Hamiltonian of first order Einstein-Cartan gravity}
\author{N. Kiriushcheva and S.V. Kuzmin}
\affiliation{Faculty of Arts and Social Science, Huron University College and Department of
Applied Mathematics, University of Western Ontario, London, Canada}
\email{nkiriush@uwo.ca, skuzmin@uwo.ca}

\begin{abstract}
Based on preliminary analysis of the Hamiltonian formulation of the first
order Einstein-Cartan action (arXiv:0902.0856 [gr-qc] and arXiv:0907.1553
[gr-qc]) we derive the Darboux coordinates, which are a unique and uniform
change of variables preserving equivalence with the original action in all
spacetime dimensions higher than two. Considerable simplification of the
Hamiltonian formulation using the Darboux coordinates, compared with direct
analysis, is explicitly demonstrated. Even an incomplete Hamiltonian analysis
in combination with known symmetries of the Einstein-Cartan action and the
equivalence of Hamiltonian and Lagrangian formulations allows us to
unambiguously conclude that the \textit{unique} \textit{gauge} invariances
generated by the first class constraints of the Einstein-Cartan action and the
corresponding Hamiltonian are \textit{translation and rotation in the tangent
space}. Diffeomorphism invariance, though a manifest invariance of the action,
is not generated by the first class constraints of the theory.

\end{abstract}
\date{\today}
\maketitle

\section{\bigskip Introduction}

In this paper we continue our search for the gauge invariance of the
Einstein-Cartan (EC) action using the Hamiltonian formulation of its first
order form, which is valid in all spacetime dimensions ($D$) higher than two
($D>2$). This investigation was started in \cite{3D, Report, Trans}. The
complete Hamiltonian analysis of the EC action when $D=3$, including the
restoration of gauge invariance, was performed in \cite{3D} because of the
simplification in calculations which appears in $D=3$. In dimensions $D>3$ the
calculations are much more involved and were not completed, though the
preliminary results were reported in \cite{Report}. The main goal of the
present paper is the derivation of the Darboux coordinates which, as we will
show, drastically simplify the calculations at the first steps of the Dirac
procedure \cite{Diracbook, Diracarticles}. We hope that additional
simplifications will also occur at the later steps and it will be possible to
complete the Hamiltonian analysis for $D>3$ and give the unique answer to the
question: what is the \textit{gauge} symmetry of the EC action, or the
symmetry which is generated by the first class constraints.

The Hamiltonian formulation of the EC action is an old and apparently solved
problem. It is claimed in many articles and included in monographs that the
canonical formulation of EC theory has already been completed and its
\textit{gauge} symmetries are Lorentz invariance and diffeomorphism (very
often only so-called \textquotedblleft spatial\textquotedblright%
\ diffeomorphism). The reasons for reconsidering this claim are the following:

A) The first order form of the EC action is a uniform formulation valid in all
$D>2$ dimensions and it seems to us very suspicious that such a drastic change
of gauge invariance in different dimensions is possible, e.g. from
Poincar\'{e} in three-dimensional case (both gauge parameters with internal
indices) \cite{Witten, 3D} to Lorentz (internal) plus diffeomorphism
(external) when $D=4$, as stated in many papers on Hamiltonian formulations of
tetrad gravity, and especially in papers and monographs on Loop Quantum
Gravity (LQG) \cite{Gambini, Thiemann}, where the spatial diffeomorphism
constraint is always assumed to be present, irrespective of what variables are
used. But it is clear even from the first steps of the Dirac procedure, as
shown in \cite{3D, Report}, that diffeomorphism (neither spatial
diffeomorphism, as in LQG, nor full spacetime diffeomorphism) \textit{cannot
be a gauge symmetry} generated by the first class constraints of the EC
action, not only in $D=3$, but in any dimension. The claim that spatial
diffeomorphism is a \textit{gauge} symmetry of tetrad gravity is the result of
a non-canonical change of variables (see Section V of \cite{Myths-2}) that was
\textquotedblleft justified\textquotedblright\ only by such \textquotedblleft
arguments\textquotedblright\ as \textquotedblleft
convenience\textquotedblright\ and a desire to accommodate the
\textquotedblleft expected\textquotedblright\ results. Of course,
diffeomorphism is an invariance of EC action (as it is manifestly generally
covariant), as well as an invariance of EC action under rotation and
translation in internal space \cite{Trautman}; in fact, many other invariances
can be found in the Lagrangian formalism by constructing differential
identities \cite{Trans}. The statement that the EC action is invariant under
internal translation when $D=3$ and is not invariant in dimensions $D>3$ is
simply wrong as this contradicts known results \cite{Trautman, Hehl-1}. The
change of \textit{gauge} symmetry from internal translation (the gauge
parameter has an internal index) to diffeomorphism (the gauge parameter is the
\textquotedblleft world\textquotedblright\ vector) does not seem to be
feasible as the first order EC action is formulated uniformly for all
dimensions ($D>2$).

Diffeomorphism is one of the invariances of the EC action but it is not a
gauge symmetry generated by the first class constraints \cite{3D, Report}. The
\textit{gauge symmetry} is a unique characteristic of a theory and in a
Hamiltonian formalism it must be uniquely derived using the Dirac procedure.
According to Dirac's conjecture \cite{Diracbook} all the first class
constraints of the Hamiltonian formulation are responsible for the
\textit{gauge} invariance and any gauge symmetry must be derivable from first
class constraints \cite{HT} using, for example, the Castellani algorithm
\cite{Castellani}. Only after that is it possible to answer the question posed
by Matschull \cite{Matschull} for $D=3$ (but which is equally well relevant in
all dimensions): \textquotedblleft what is a gauge symmetry and what is
not\textquotedblright. We are not aware of such a derivation for the
Hamiltonian formulation of the EC action, similar to the Einstein-Hilbert,
metric, action \cite{KKRV, Myths}, where full spacetime diffeomorphism is
indeed gauge symmetry. Some arguments that so-called spatial diffeomorphism is
a gauge symmetry have been made; but this is not even a symmetry of the EC
action which is invariant under the full spacetime diffeomorphism, not under
its spatial part separately.

B) It was shown some time ago that the EC action is invariant under
\textquotedblleft translations and rotations in the tangent
spaces\textquotedblright\ \cite{Trautman} with an algebra of generators that
has \textquotedblleft a more general group structure than the original
Poincar\'{e} group\textquotedblright\ \cite{Hehl-1}. It differs from the
original Poincar\'{e} group for $D>3$ only by having a non-zero commutation
relation between two translational generators \cite{Trautman, Hehl-1}.
Moreover, the explicit form of the transformations of fields were given with
two parameters that correspond to internal translation and rotation (see e.g.
\cite{Hehl-1, Leclerc}). These results contradict the statement that the EC
Lagrangian is not invariant under translation. Recently (without using the
Hamiltonian formalism) translational and rotational invariances of the first
order EC action were derived in \cite{Trans} by constructing the simplest
differential identities from the Euler derivatives which follow from the EC
Lagrangian. To find a gauge invariance of the EC action, a Hamiltonian
analysis is needed. However, the choice of what to call a gauge invariance is
often based on different arguments, i.e. according to \cite{Hehl-2}
\textquotedblleft it is partly possible (and physically more plausible) to
unify the two local gauge groups - Poincar\'{e} on the frames and general
covariance\textquotedblright. Is it possible to have the \textit{gauge} group
that unify Poincar\'{e} and diffeomorphism? Lagrangian and Hamiltonian
formalisms are equivalent; and in the Hamiltonian formalism, the
\textit{gauge} invariance must be derived from first class constraints
\cite{HT}, not to be defined arbitrarily. The number of constraints not only
fixes the number of gauge parameters (which equals to the number of primary
first class constraints) and their tensorial character but it also defines the
number of degrees of freedom (found from the number of all constraints)
\cite{HTZ}. Thus, internal translation and diffeomorphism cannot
simultaneously be \textit{gauge} group, as the number of constraints needed to
accommodate both symmetries lead to a negative number of degrees of freedom,
which is physically not plausible. To find a unique gauge symmetry from all
possible invariances of an action cannot be done unambiguously without a
Hamiltonian analysis. We do not rely on geometrical, or physical, or any other
plausible argument; and our \textit{goal is to reveal the gauge symmetry of
the EC theory using the Hamiltonian method.}

C) The Hamiltonian analysis of the first order form of the EC action is
specialized to some particular dimension, e.g. \cite{Plebanski, Ashtekar,
Witten, Peldan, Banados}. However, such formulations might either destroy or
miss some general features of the original action, which are valid in all
dimensions (except the special case $D=2$). For example, in constructing
Darboux coordinates (the main subject of our article), it would be artificial
to introduce such variables for each dimension separately - they have to be
common for all dimensions as the original EC action is.

Our method of finding Darboux variables is valid in all dimensions $D>2$. It
is not a purely mathematical interest to find the most general formulation,
but it has a practical reason: the EC action is formulated in all dimensions,
and so the correct methods have to produce meaningful results in all
dimensions simultaneously. This will guarantee that nothing is missing or
misinterpreted in the physically important four-dimensional case. Examples of
formulations that were designed for only particular dimensions are: when $D=3$
the treatment of the EC action based on similarities (but not equivalence
\cite{Matschull}) with the Chern-Simons action \cite{Witten}; the construction
of Darboux coordinates by Ba\~{n}ados and Contreras \cite{Banados} that works
only in the $D=4$ case and allows for neither consideration of the $D=3$ limit
nor dimensions higher than four. To have the correct Hamiltonian formulation
of the EC action, and to find its unique gauge invariance in the physically
interesting $D=4$ case, we have to perform the analysis using an approach
valid in all dimensions. An important property of using a formulation valid in
all dimensions is the possibility to check the $D=3$ limit at all stages of
the calculations. The Hamiltonian formulation in the $D=3$ case is simple,
gives the consistent result and a simple Lie algebra of Poisson brackets (PBs)
among the first class constraints \cite{3D}. In higher dimensions we can
expect that some modifications of the Poincar\'{e} algebra will appear (we
argued in \cite{Report} that the only possible modification is the non-zero PB
among two translational constraints, which is exactly what happens in
\cite{Trautman, Hehl-1}); but such modifications must disappear in the $D=3$
limit. The calculation of constraints and their PB algebra in the Hamiltonian
formulation of the EC action are still quite involved even after drastic
simplification due to introduction of Darboux coordinates, and so the
possibility of checking the consistency of the results by considering the
$D=3$ limit, at all stages of calculations, is extremely important. Thus,
\textit{we do not specialize our analysis to a particular dimension.}

The construction of Darboux coordinates which are uniform for all dimensions
and simplify the Hamiltonian analysis is the main goal of this article. 

The paper is organized as follows. In the next Section (II) we establish
notation and provide arguments to support our expectation that Darboux
coordinates exist. In Section III, based on the result of the direct
Hamiltonian analysis \cite{Report}, we derive Darboux coordinates. In Section
IV, we show that introduction of Darboux coordinates allows one to perform the
Lagrangian or Hamiltonian reduction in a much simpler manner than the
Hamiltonian reduction in the direct Hamiltonian approach of \cite{Report}, and
to attack the most involved calculations: finding PBs among secondary first
class constraints (or equivalently, to prove the closure of the Dirac
procedure) which is needed to find gauge transformations of the EC action in
the Hamiltonian formalism. These calculations are briefly outlined. In
particular, using Dirac brackets, we demonstrate that there is a strong
indication that in all dimensions the PB between translational and rotational
constraints are the same and coincide with the corresponding part of the
Poincar\'{e} algebra known for the $D=3$ case (the same conclusion was made in
\cite{Report} using the Castellani algorithm \cite{Castellani}). In the last
Section (V) the results are summarized and conclusion about \textit{gauge}
invariance of the EC action is made. The properties of some combinations of
fields that considerably simplify the calculations are collected in Appendix
A. In Appendix B the solution of the equation that arises in the course of the
Lagrangian/Hamiltonian reduction is given.

\section{Notation and expectations}

In \cite{Report} we considered the Hamiltonian formulation of the
Einstein-Cartan action by direct application of the Dirac procedure to its
first order form \cite{Sch, CNP}%

\begin{equation}
I_{EC}=-\int d^{D}x\ e\left(  e^{\mu\left(  \alpha\right)  }e^{\nu\left(
\beta\right)  }-e^{\nu\left(  \alpha\right)  }e^{\mu\left(  \beta\right)
}\right)  \left(  \omega_{\nu\left(  \alpha\beta\right)  ,\mu}+\omega
_{\mu\left(  \alpha\gamma\right)  }\omega_{\nu~~\beta)}^{~(\gamma}\right)
\label{eqnD1}%
\end{equation}
where the covariant N-beins $e_{\gamma\left(  \rho\right)  }$ and the
connections $\omega_{\nu\left(  \alpha\beta\right)  }$ ($\omega_{\nu\left(
\alpha\beta\right)  }=-\omega_{\nu\left(  \beta\alpha\right)  }$) are treated
as independent fields in all spacetime dimensions ($D>2$), and $e=\det\left(
{e_{\gamma\left(  \rho\right)  }}\right)  $.\footnote{Usually variables
$e_{\gamma\left(  \rho\right)  }$ and $\omega_{\nu\left(  \alpha\beta\right)
}$ are named tetrads and spin connections, but such names are specialized for
$D=4$. As we consider the Hamiltonian formulation in any dimension ($D>2$), we
will call $e_{\gamma\left(  \rho\right)  }$ and $\omega_{\nu\left(
\alpha\beta\right)  }$ N-beins and connections, respectively.} Greek letters
indicate covariant indices $\alpha=0,1,2,...,\left(  D-1\right)  $. Indices in
brackets $(...)$ denote the internal (\textquotedblleft
Lorentz\textquotedblright) indices, whereas indices without brackets are
external or \textquotedblleft world\textquotedblright\ indices. Internal and
external indices are raised and lowered by the Minkowski tensor $\tilde{\eta
}_{\alpha\beta}=\left(  -,+,+,...\right)  $ and the metric tensor $g_{\mu\nu
}=e_{\mu\left(  \alpha\right)  }e_{\nu}^{\left(  \alpha\right)  }$,
respectively (we use a tilde for any combination with only internal indices
and do not use brackets in such cases, except to indicate antisymmetrization
in pairs of indices). N-beins are invertible: $e^{\mu\left(  \alpha\right)
}e_{\mu\left(  \beta\right)  }=\tilde{\delta}_{\beta}^{\alpha}$,
$e^{\mu\left(  \alpha\right)  }e_{\nu\left(  \alpha\right)  }=\delta_{\nu
}^{\mu}$.

The Lagrangian density of (\ref{eqnD1}), after an integration by parts, can be
written in the following form%

\begin{equation}
L_{EC}\left(  e_{\mu\left(  \alpha\right)  },\omega_{\mu\left(  \alpha
\beta\right)  }\right)  =eB^{\gamma\left(  \rho\right)  \mu\left(
\alpha\right)  \nu\left(  \beta\right)  }e_{\gamma\left(  \rho\right)  ,\mu
}\omega_{\nu\left(  \alpha\beta\right)  }-eA^{\mu\left(  \alpha\right)
\nu\left(  \beta\right)  }\omega_{\mu\left(  \alpha\gamma\right)  }\omega
_{\nu~~\beta)}^{~(\gamma} \label{eqnD2}%
\end{equation}
where the functions $A^{\mu\left(  \alpha\right)  \nu\left(  \beta\right)  }$
and $B^{\gamma\left(  \rho\right)  \mu\left(  \alpha\right)  \nu\left(
\beta\right)  }$ are defined as%

\begin{equation}
A^{\mu\left(  \alpha\right)  \nu\left(  \beta\right)  }=e^{\mu\left(
\alpha\right)  }e^{\nu\left(  \beta\right)  }-e^{\nu\left(  \alpha\right)
}e^{\mu\left(  \beta\right)  },\text{ \ \ }\frac{\delta}{\delta e_{\gamma
\left(  \rho\right)  }}\left(  eA^{\mu\left(  \alpha\right)  \nu\left(
\beta\right)  }\right)  =eB^{\gamma\left(  \rho\right)  \mu\left(
\alpha\right)  \nu\left(  \beta\right)  }.\label{eqnD3}%
\end{equation}

The properties of the function $A^{\mu\left(  \alpha\right)  \nu\left(
\beta\right)  }$ and $B^{\gamma\left(  \rho\right)  \mu\left(  \alpha\right)
\nu\left(  \beta\right)  }$ and their further generations that considerably
simplify the calculations are collected in Appendix A.

For the Hamiltonian formulation, where we have to separate spatial and
temporal \textit{indices }(not separating spacetime itself into space and
time),\footnote{If one writes, for example, equations of motion of a covariant
theory in components the covariance is not lost, though it is not manifest.
The common statement as in \cite{Pullin-1}: \textquotedblleft Unfortunately,
the canonical treatment breaks the symmetry between space and time in general
relativity and the resulting algebra of constraints is not the algebra of four
diffeomorphism\textquotedblright\ is groundless. In the Hamiltonian
formulation of General Relativity the covariance is not manifest, but it is
not broken as the gauge symmetry of the Einstein-Hilbert action,
diffeomorphism, is recovered in a manifestly covariant form for the second
order \cite{KKRV, Myths} and the first order \cite{Myths-2} formulations.} we
use $0$ for an external \textquotedblleft time\textquotedblright\ index (and
$\left(  0\right)  $ for an internal \textquotedblleft time\textquotedblright%
\ index) and Latin letters for \textquotedblleft spatial\textquotedblright%
\ external indices $k=1,2,...,\left(  D-1\right)  $ ($\left(  k\right)  $ for
\textquotedblleft spatial\textquotedblright\ internal indices).

In Progress Report \cite{Report} (the references to equations from
\cite{Report} are indicated as Eq. (R\#)) we demonstrated that after
performing the Hamiltonian reduction (i.e. elimination of part of the
variables by solving the second class constraints) in all dimensions, the
canonical part of the total Hamiltonian is a linear combination of secondary
constraints (called \textquotedblleft rotational\textquotedblright%
\ $\chi^{0\left(  \alpha\beta\right)  }$ and \textquotedblleft
translational\textquotedblright\ $\chi^{0\left(  \sigma\right)  }$
constraints, see Eq. (R152))%

\begin{equation}
H_{reduced}\left(  e_{\mu\left(  \rho\right)  },\pi^{\mu\left(  \rho\right)
},\omega_{0\left(  \alpha\beta\right)  },\Pi^{0\left(  \alpha\beta\right)
}\right)  =\pi^{0\left(  \rho\right)  }\dot{e}_{0\left(  \rho\right)  }%
+\Pi^{0\left(  \alpha\beta\right)  }\dot{\omega}_{0\left(  \alpha\beta\right)
}+H_{c}~,\label{eqnD4}%
\end{equation}
where the canonical Hamiltonian (up to a total spatial derivative) is%

\begin{equation}
H_{c}=-\omega_{0\left(  \alpha\beta\right)  }\chi^{0\left(  \alpha
\beta\right)  }\left(  e_{\mu\left(  \rho\right)  },\pi^{k\left(  \rho\right)
}\right)  -e_{0\left(  \sigma\right)  }\chi^{0\left(  \sigma\right)  }\left(
e_{\mu\left(  \rho\right)  },\pi^{k\left(  \rho\right)  }\right)  .
\label{eqnD5}%
\end{equation}

For $D=4$ the reduced Hamiltonian with the same set of canonical variables
(\ref{eqnD4}) was obtained in \cite{CNP}.

All PBs among primary ($\pi^{0\left(  \rho\right)  },\Pi^{0\left(  \alpha
\beta\right)  }$) and among primary and secondary constraints ($\chi^{0\left(
\alpha\beta\right)  },\chi^{0\left(  \sigma\right)  }$) are zero and the PB
between two rotational constraints in all dimensions is%

\begin{equation}
\left\{  \chi^{0\left(  \alpha\beta\right)  },\chi^{0\left(  \mu\nu\right)
}\right\}  _{D>2}=\frac{1}{2}\tilde{\eta}^{\beta\mu}\chi^{0\left(  \alpha
\nu\right)  }-\frac{1}{2}\tilde{\eta}^{\alpha\mu}\chi^{0\left(  \beta
\nu\right)  }+\frac{1}{2}\tilde{\eta}^{\beta\nu}\chi^{0\left(  \mu
\alpha\right)  }-\frac{1}{2}\tilde{\eta}^{\alpha\nu}\chi^{0\left(  \mu
\beta\right)  }, \label{eqnD6}%
\end{equation}
which corresponds to Lorentz rotation in the tangent space. It is necessary to
find the remaining PBs:%

\begin{equation}
\left\{  \chi^{0\left(  \alpha\beta\right)  },\chi^{0\left(  \rho\right)
}\right\}  _{D>3}=?\text{ \ ,\ \ \ \ } \label{eqnD7}%
\end{equation}

\begin{equation}
\left\{  \chi^{0\left(  \rho\right)  },\chi^{0\left(  \gamma\right)
}\right\}  _{D>3}=?\text{ \ .}\label{eqnD7a}%
\end{equation}
In the $D=3$ case, the calculation of (\ref{eqnD7})-(\ref{eqnD7a}) is simple
\cite{3D} and leads to:%

\begin{equation}
\left\{  \chi^{0\left(  \alpha\beta\right)  },\chi^{0\left(  \rho\right)
}\right\}  _{D=3}=\frac{1}{2}\tilde{\eta}^{\beta\rho}\chi^{0\left(
\alpha\right)  }-\frac{1}{2}\tilde{\eta}^{\alpha\rho}\chi^{0\left(
\beta\right)  },\text{ \ \ } \label{eqnD8}%
\end{equation}

\begin{equation}
\left\{  \chi^{0\left(  \rho\right)  },\chi^{0\left(  \gamma\right)
}\right\}  _{D=3}=0, \label{eqnD8a}%
\end{equation}
i.e. the PB algebra of secondary constraints (\ref{eqnD6}), (\ref{eqnD8}) and
(\ref{eqnD8a}) is a true Poincar\'{e} algebra; and a complete set of first
class constraints when using the Castellani algorithm leads to the rotational
and translational invariance in tangent space \cite{3D}. In higher dimensions,
even knowledge of only the primary constraints is enough to conclude that it
is impossible to have diffeomorphism invariance following from the first class
constraints and the \textit{gauge} parameters must possess internal indices,
i.e. they lead to rotation and translation\ in the internal space. Whether it
is a true Poincar\'{e} algebra or a modified Poincar\'{e}, can only be found
after the PBs (\ref{eqnD7})-(\ref{eqnD7a}) are calculated.

In higher dimensions, calculation of the PBs of (\ref{eqnD7})-(\ref{eqnD7a})
is very laborious because of the complexity of the constraints \cite{Report}.
Nevertheless, these PBs are needed to prove the closure of the Dirac
procedure, and to find the transformations that are produced by the first
class constraints, i.e. to answer the question (in the Hamiltonian formalism)
of which symmetry (from an infinite set of symmetries of the EC action
\cite{Trans}) is the \textit{gauge} symmetry of the EC action.

In the conclusion of \cite{Report} we discussed possible modifications of the
algebra of the PBs in dimensions $D>3$, based on the assumption that despite a
more complicated form of constraints the algebra of secondary constraints
remains Poincar\'{e}, as in the $D=3$ case, or is modified Poincar\'{e}
algebra. We showed that in such cases, the Lagrangian corresponding to the
reduced Hamiltonian (\ref{eqnD5}) remains invariant. This Lagrangian can be
obtained by performing the inverse Legendre transformations and it gives us
just a different first order formulation of the original EC theory (Eq. (R162))%

\[
L_{reduced}\left(  e_{\mu\left(  \alpha\right)  },\pi^{k\left(  \rho\right)
},\omega_{0\left(  \alpha\beta\right)  }\right)  =
\]

\begin{equation}
\pi^{k\left(  \rho\right)  }\dot{e}_{k\left(  \rho\right)  }+\omega_{0\left(
\alpha\beta\right)  }\chi^{0\left(  \alpha\beta\right)  }\left(  e_{\mu\left(
\rho\right)  },\pi^{k\left(  \rho\right)  }\right)  +e_{0\left(
\sigma\right)  }\chi^{0\left(  \sigma\right)  }\left(  e_{\mu\left(
\rho\right)  },\pi^{k\left(  \rho\right)  }\right)  .\label{eqnD9}%
\end{equation}
In the Lagrangian formalism $\pi^{k\left(  \rho\right)  }$, as well as
$\omega_{0\left(  \alpha\beta\right)  }$, are just auxiliary variables that
can be eliminated using their equations of motion (exactly as $\omega
_{\mu\left(  \alpha\beta\right)  }$ can be eliminated in (\ref{eqnD1})) that
leads back to the second order EC action. The Lagrangian (\ref{eqnD9}) gives
the first order formulation of the EC action that differs in field content
from (\ref{eqnD1}) but they both are equivalent to second order form of the EC
action (after elemination of auxiliary fields). However, the first order form
(\ref{eqnD9}) leads directly to the Hamiltonian.

We can say that the following operations were performed:%

\begin{equation}
L_{EC}\rightarrow H\underset{\text{Hamiltonian/Dirac reduction}%
}{\Longrightarrow}H_{reduced}\text{ (Eq. }\left(  \ref{eqnD4}\right)
\text{)}\rightarrow L_{reduced}\text{ (Eq. }\left(  \ref{eqnD9}\right)
\text{).} \label{eqnD10}%
\end{equation}

This suggests (because the Hamiltonian and Lagrangian formalisms must lead to
the same result, of course, if the reductions are performed correctly
\cite{AOP}) that such a reduced Lagrangian should be possible to obtain
directly from the EC action, so there should be some transformations which
give Darboux coordinates that can simplify the calculations of $H_{reduced}$
compared to the direct calculations \cite{Report} performed for (\ref{eqnD2}).
Perhaps, it can simplify calculations of the remaining PBs among secondary
constraints (\ref{eqnD7})-(\ref{eqnD7a}) and the corresponding gauge
transformations. In other words, we are looking for Darboux coordinates such that%

\[
L_{EC}\underset{\text{Darboux coordinates}}{\Longrightarrow}L_{EC\left(
D\right)  }\underset{\text{Lagrangian reduction}}{\Longrightarrow}L_{reduced}%
\]
with%

\begin{equation}
L_{reduced}=L_{reduced}\text{ (Eq. }\left(  \ref{eqnD9}\right)  \text{)}%
\rightarrow H_{reduced}\text{ (Eq. }\left(  \ref{eqnD4}\right)  \text{).}%
\label{eqnD11}%
\end{equation}

In the next Section, based on the result of the direct Hamiltonian analysis
\cite{Report}, we derive the following Darboux coordinates for spatial
connections (the temporal connections $\omega_{0\left(  \alpha\beta\right)
},$ as well as the basic variables $e_{\gamma\left(  \rho\right)  }$, N-beins,
remain unaltered)%

\begin{equation}
\omega_{m\left(  \alpha\beta\right)  }=N_{m\left(  \alpha\beta\right)
0n\left(  \sigma\right)  }F^{n\left(  \sigma\right)  }+e_{p\left(
\alpha\right)  }e_{q\left(  \beta\right)  }\hat{\Sigma}_{m}^{\ \left(
pq\right)  }=\omega_{m\left(  \alpha\beta\right)  }\left(  F\right)
+\omega_{m\left(  \alpha\beta\right)  }\left(  \hat{\Sigma}\right)
,\label{eqnD12}%
\end{equation}
where $N_{m\left(  \alpha\beta\right)  0n\left(  \sigma\right)  }$ is a
non-linear, algebraic (without derivatives) combination of N-beins which is
antisymmetric in $\alpha\beta$ whose explicit form is given by (\ref{eqnD13}).
The field $\hat{\Sigma}_{m}^{\ \left(  pq\right)  }$ is antisymmetric
($\hat{\Sigma}_{m}^{\ \left(  pq\right)  }=-\hat{\Sigma}_{m}^{\ \left(
qp\right)  }$) and traceless ($\hat{\Sigma}_{m}^{\ \left(  mq\right)  }%
=\hat{\Sigma}_{m}^{\ \left(  pm\right)  }=0$) with all indices being external
(\textquotedblleft world\textquotedblright) and spatial. (Here and below we
will use \textquotedblleft hat\textquotedblright\ for combinations with only
external indices and brackets used to indicate antisymmetrization in pairs of
external indices.) The transformation (\ref{eqnD12}) is invertible, valid in
all dimensions ($D>2$) and preserves the $D=3$ limit \cite{3D}. Note that the
number of components of $\hat{\Sigma}_{m}^{\ \left(  pq\right)  }$ plus
$F^{n\left(  \sigma\right)  }$ is the same as of $\omega_{m\left(  \alpha
\beta\right)  }$ in all dimensions (the number of independent components of a
field, using the notation of \cite{GT}, is defined by $\left[  ...\right]  $):%

\begin{equation}
\left[  \omega_{m\left(  \alpha\beta\right)  }\right]  =\frac{1}{2}D\left(
D-1\right)  \left(  D-1\right)  , \label{eqnD15}%
\end{equation}

\begin{equation}
\left[  F^{n\left(  \sigma\right)  }\right]  =D\left(  D-1\right)  ,
\label{eqnD16}%
\end{equation}

\begin{equation}
\left[  \hat{\Sigma}_{m}^{\ \left(  pq\right)  }\right]  =\frac{1}{2}D\left(
D-1\right)  \left(  D-3\right)  \label{eqnD17}%
\end{equation}
that gives%

\begin{equation}
\left[  \omega_{m\left(  \alpha\beta\right)  }\right]  =\left[  \pi^{n\left(
\sigma\right)  }\right]  +\left[  \hat{\Sigma}_{m}^{\ \left(  pq\right)
}\right]  . \label{eqnD18}%
\end{equation}

In the discussion of Darboux coordinates specialized to $D=4$ case appearing
in \cite{Banados}, a different field is introduced, $\hat{\lambda}_{km}%
=\hat{\lambda}_{mk}$, instead of our $\hat{\Sigma}_{m}^{\ \left(  pq\right)
}$. The number of components is $\left[  \hat{\lambda}_{km}\right]  =$
$\frac{1}{2}D\left(  D-1\right)  $ which gives the correct balance of fields
(see (\ref{eqnD18})) only in the $D=4$ case (as in this dimension $\left[
\hat{\lambda}_{km}\right]  =\left[  \hat{\Sigma}_{m}^{\ \left(  pq\right)
}\right]  =6$), but supports neither a $D=3$ limit nor a generalization to
higher dimensions. The uniform description of the EC action in all dimensions
is broken by such coordinates.

In Section IV, we show that using the transformation (\ref{eqnD12}) not only
diagonalizes the \textquotedblleft kinetic\textquotedblright\ part of the
Lagrangian (terms with temporal derivatives of N-beins), as a consequence of
the following properties%

\begin{equation}
eB^{k\left(  \rho\right)  0\left(  \alpha\right)  m\left(  \beta\right)
}\omega_{m\left(  \alpha\beta\right)  }\left(  F\right)  =F^{k\left(
\rho\right)  },\text{ \ \ \ }B^{k\left(  \rho\right)  0\left(  \alpha\right)
m\left(  \beta\right)  }\omega_{m\left(  \alpha\beta\right)  }\left(
\hat{\Sigma}\right)  =0,\label{eqnD19}%
\end{equation}
but also provides a separation of variables that allows one to perform the
Lagrangian or Hamiltonian reduction in a much simpler manner than in the
direct Hamiltonian approach of \cite{Report} (i.e. to eliminate the field
$\hat{\Sigma}_{m}^{\ \left(  pq\right)  }$, which corresponds to solution of
the secondary second class constraints in the Hamiltonian analysis). After
elimination of $\hat{\Sigma}_{m}^{\ \left(  pq\right)  }$, the Hamiltonian
(\ref{eqnD4})-(\ref{eqnD5}), where only first class constraints are present,
can be just read off from the reduced first order Lagrangian (\ref{eqnD9}).
The Hamiltonian and the first class constraints that were obtained after long
and cumbersome calculations in the direct approach of \cite{Report} which
started from (\ref{eqnD2}) can be found almost immediately when Darboux
coordinates are introduced. We show that simplifications due to introduction
of the Darboux coordinates allows us to attack the most involved calculations
in the direct approach: finding the remaining PBs (\ref{eqnD7})-(\ref{eqnD7a})
among secondary constraints (or equivalently, to prove the closure of the
Dirac procedure) which is needed to find gauge transformations of the EC
action in the Hamiltonian formalism. These calculations will be briefly
outlined. In particular, we demonstrate that there is a strong indication that
in all dimensions the PB between translational and rotational constraints are
the same and coincide with the corresponding part of the Poincar\'{e} algebra
(\ref{eqnD8}) known for the $D=3$ case \cite{3D} (the same conclusion was made
in \cite{Report} using the Castellani algorithm):%

\begin{equation}
\left\{  \chi^{0\left(  \alpha\beta\right)  },\chi^{0\left(  \rho\right)
}\right\}  _{D>2}=\frac{1}{2}\tilde{\eta}^{\beta\rho}\chi^{0\left(
\alpha\right)  }-\frac{1}{2}\tilde{\eta}^{\alpha\rho}\chi^{0\left(
\beta\right)  }. \label{eqnD20}%
\end{equation}

\section{Derivation of Darboux coordinates using a preliminary Hamiltonian
analysis of the Einstein-Cartan action}

Direct application of the Dirac procedure to the first order formulation of
Einstein-Cartan action without specialization to a particular dimension was
discussed in \cite{Report} where after performing the Hamiltonian reduction
(that is, elimination of second class constraints) the total Hamiltonian
(\ref{eqnD4})-(\ref{eqnD5}) was obtained. However, these calculations are
extremely laborious and on the last stage (closure of the Dirac procedure)
become almost unmanageable with the exception of the $D=3$ case \cite{3D}. But
this preliminary Hamiltonian analysis is indispensable because it allows us to
find variables at the Lagrangian level that drastically simplify the first
steps of the calculations. At the Lagrangian level, only the invertability of
the change of variables is usually checked; but it might happen that the
change of variables even being invertible is not canonical in the Hamiltonian
formulation. That is why, in finding new variables it is important to rely on
the Hamiltonian analysis, especially when working with systems which have
first and second class constraints. We perform a classification of fields
according to their relation to the constraints arising in the Hamiltonian
formulation. This specific role of different fields can be used also at the
Lagrangian level and, in particular, allows one to find Darboux coordinates
that preserve equivalence with the original action and are helpful in reducing
the amount of calculation that must be done. We will use the classification of
fields that corresponds to the classification of constraints. \textit{Primary}
constraints (using a notion introduced by Bergmann), especially \textit{first
class} (a notion introduced by Dirac), play the most important role in the
Hamiltonian formulation and define the tensorial character of the gauge
parameters (see Section 5 of \cite{Myths-2}), so we call the variables
\textquotedblleft\textit{primary variables}\textquotedblright\ if in the
Hamiltonian formulation the corresponding momenta enter the primary first
class constraints. For the first order Einstein-Cartan action the primary
variables are $e_{0\left(  \alpha\right)  }$ and $\omega_{0\left(  \alpha
\beta\right)  }$ \cite{Report}. \textit{Second} \textit{class} constraints
(using Dirac's classification) irrespective of their generation (primary,
secondary, etc.) can be solved for pairs of phase-space variables (this is the
Hamiltonian reduction). We call such variables \textquotedblleft\textit{second
class variables}\textquotedblright, i.e. variables that at the Hamiltonian
level can be eliminated. In the case of the first order Einstein-Cartan action
second class variables are $\omega_{k\left(  \alpha\beta\right)  }$
\cite{Report}. Primary and second class variables of the first order EC action
could be already identified from the results of the first gauge-free (without
\textit{a priori} choice of a particular gauge) Hamiltonian formulation for
$D=4$ \cite{CNP}. The importance of careful preliminary analysis before doing
the Lagrange reduction was emphasized in \cite{GGT}: \textquotedblleft it
seems important to develop reduction procedure within Lagrangian formulation -
in a sense similar to the Dirac procedure in the Hamiltonian formulation -
that may allow one to reveal the hidden structure of the Euler-Lagrange
equations of motion in a constructive manner\textquotedblright.

The above classification is crucial because the equivalence of the Lagrangian
and Hamiltonian methods dictates that if changes involving primary variables
are very restrictive at the Hamiltonian level \cite{Myths-2} then the same
must be true also at the Lagrangian level. An arbitrary change of variables
can lead to the loss of equivalence of two formulations even for changes which
are invertible, that is the sufficient condition only for nonsingular systems.
The second class fields can be eliminated and there is more freedom to
redefine them; but this redefinition has to be such that their elimination
does not modify the PBs for the remaining fields, if this is what happens in
the Hamiltonian formalism. This imposes some restrictions; and even in this
case, the invertability of transformations is only a necessary condition. In
particular, in the construction of the Darboux coordinates for second class
variables (which can be a complicated expression) they have to be independent
of the primary variables, i.e. their variation with respect to primary
variables must be zero so as to preserve the original independence of primary
and second class variables. From the above arguments it is clear that the
Hamiltonian analysis is indispensable if one wants to modify the Lagrangian
while keeping its equivalence with the original one (for the gauge-invariant
systems). For example, the original independence of primary and second class
variables, $\frac{\delta\omega_{k\left(  \alpha\beta\right)  }}{\delta
e_{0\left(  \rho\right)  }}=\frac{\delta\omega_{k\left(  \alpha\beta\right)
}}{\delta\omega_{0\left(  \rho\sigma\right)  }}=0$, should be preserved even
after a change of variables in which Darboux coordinates are introduced for
$\omega_{k\left(  \alpha\beta\right)  }$.

In this Section, we describe the construction of Darboux coordinates for the
EC action and also illustrate the general points mentioned above. Our goal is
to find the Darboux coordinates for the second class fields that simplify the
Hamiltonian analysis. In \cite{Report}, in the course of the Hamiltonian
reduction, all $\omega_{k\left(  \alpha\beta\right)  }$ were eliminated by
solving the second class constraints: one part by solving the primary
constraints and another part that involved primary and secondary constraints.
So, it would be preferable to find such a representation of $\omega_{k\left(
\alpha\beta\right)  }$ that separate its components into exactly two classes
of variables, as in the Hamiltonian they were mixed leading to quite long
calculations. We want to decouple them, i.e. we have to find such a
transformation of an \textquotedblleft auxiliary\textquotedblright, second
class,\ field $\omega_{k\left(  \alpha\beta\right)  }$ (Darboux coordinates)
that diagonalizes the \textquotedblleft kinetic part\textquotedblright\ and
separate variables that can be eliminated by a Lagrangian reduction.\ Note
that such a separation automatically appears in the $D=3$ case \cite{3D}.

The direct Hamiltonian analysis of the first order EC action (\ref{eqnD1})
starts by introducing momenta conjugate to all independent variables. In
(\ref{eqnD2}) the only term that has \textquotedblleft
velocities\textquotedblright\ is%

\begin{equation}
L\left(  e_{\gamma\left(  \rho\right)  ,0}\right)  =eB^{\gamma\left(
\rho\right)  0\left(  \alpha\right)  \nu\left(  \beta\right)  }e_{\gamma
\left(  \rho\right)  ,0}\omega_{\nu\left(  \alpha\beta\right)  }
\label{eqnD21}%
\end{equation}
and so the momenta corresponding to $e_{\gamma\left(  \rho\right)  }$ are
defined as%

\begin{equation}
\pi^{\gamma\left(  \rho\right)  }=\frac{\delta L}{\delta e_{\gamma\left(
\rho\right)  ,0}}=eB^{\gamma\left(  \rho\right)  0\left(  \alpha\right)
\nu\left(  \beta\right)  }\omega_{\nu\left(  \alpha\beta\right)  }.
\label{eqnD22}%
\end{equation}
The only non-zero contributions (based on antisymmetry properties of
$B^{\gamma\left(  \rho\right)  \mu\left(  \alpha\right)  \nu\left(
\beta\right)  }$ (see Appendix A and also \cite{Trans, Report})) are%

\begin{equation}
\pi^{k\left(  \rho\right)  }=eB^{k\left(  \rho\right)  0\left(  \alpha\right)
m\left(  \beta\right)  }\omega_{m\left(  \alpha\beta\right)  } \label{eqnD23}%
\end{equation}
or upon separating $\pi^{k\left(  \rho\right)  }$ and $\omega_{m\left(
\alpha\beta\right)  }$ into \textquotedblleft space\textquotedblright\ and
\textquotedblleft time\textquotedblright\ components%

\begin{equation}
\pi^{k\left(  n\right)  }=eB^{k\left(  n\right)  0\left(  p\right)  m\left(
q\right)  }\omega_{m\left(  pq\right)  }+2eB^{k\left(  n\right)  0\left(
q\right)  m\left(  0\right)  }\omega_{m\left(  q0\right)  }~,\label{eqnD24}%
\end{equation}

\begin{equation}
\pi^{k\left(  0\right)  }=eB^{k\left(  0\right)  0\left(  p\right)  m\left(
q\right)  }\omega_{m\left(  pq\right)  }. \label{eqnD25}%
\end{equation}

Equations (\ref{eqnD24}) and (\ref{eqnD25}) lead to two primary second class constraints%

\begin{equation}
\phi^{k\left(  n\right)  }=\pi^{k\left(  n\right)  }-eB^{k\left(  n\right)
0\left(  p\right)  m\left(  q\right)  }\omega_{m\left(  pq\right)
}-2eB^{k\left(  n\right)  0\left(  q\right)  m\left(  0\right)  }%
\omega_{m\left(  q0\right)  }\approx0,\label{eqnD24a}%
\end{equation}

\begin{equation}
\phi^{k\left(  0\right)  }=\pi^{k\left(  0\right)  }-eB^{k\left(  0\right)
0\left(  p\right)  m\left(  q\right)  }\omega_{m\left(  pq\right)  }%
\approx0.\label{eqnD25a}%
\end{equation}

After introducing the following notation (see \cite{Report})%

\[
\gamma^{k\left(  n\right)  }\equiv e^{k\left(  n\right)  }-\frac{e^{k\left(
0\right)  }e^{0\left(  n\right)  }}{e^{0\left(  0\right)  }}~,\quad
\gamma^{k\left(  n\right)  }e_{p\left(  n\right)  }=\delta_{p}^{k}%
~,\quad\gamma^{k\left(  n\right)  }e_{k\left(  m\right)  }=\tilde{\delta}%
_{m}^{n}~,
\]

\[
E^{k\left(  p\right)  m\left(  q\right)  }\equiv\gamma^{k\left(  p\right)
}\gamma^{m\left(  q\right)  }-\gamma^{k\left(  q\right)  }\gamma^{m\left(
p\right)  },\quad I_{m\left(  q\right)  n\left(  r\right)  }\equiv\frac
{1}{D-2}e_{m\left(  q\right)  }e_{n\left(  r\right)  }-e_{m\left(  r\right)
}e_{n\left(  q\right)  }~,
\]

\[
E^{k\left(  p\right)  m\left(  q\right)  }I_{m\left(  q\right)  n\left(
r\right)  }=\delta_{n}^{k}\tilde{\delta}_{r}^{p}~,
\]
equation (\ref{eqnD24a}) (because it is a second class constraint in the
Hamiltonian analysis) can be solved for $\omega_{k\left(  q0\right)  }$ (see
Eq. (R46))%

\begin{equation}
\omega_{k\left(  q0\right)  }=-\frac{1}{2ee^{0\left(  0\right)  }}I_{k\left(
q\right)  m\left(  p\right)  }\pi^{m\left(  p\right)  }-\frac{e^{0\left(
p\right)  }}{2e^{0\left(  0\right)  }}I_{k\left(  q\right)  m\left(  p\right)
}E^{m\left(  a\right)  n\left(  b\right)  }\omega_{n\left(  ab\right)  }%
+\frac{e^{0\left(  a\right)  }}{e^{0\left(  0\right)  }}\omega_{k\left(
aq\right)  } \label{eqnD30}%
\end{equation}
and (\ref{eqnD25a}) can be written in the following form%

\begin{equation}
\pi^{k\left(  0\right)  }=-ee^{0\left(  0\right)  }E^{k\left(  p\right)
m\left(  q\right)  }\omega_{m\left(  pq\right)  }. \label{eqnD31}%
\end{equation}

When $D=3$ (and only when $D=3$) equation (\ref{eqnD31}) can be solved for
$\omega_{m\left(  pq\right)  }$ and in equation (\ref{eqnD24a}) terms
proportional to the connections $\omega_{m\left(  pq\right)  }$ (with all
\textquotedblleft space\textquotedblright\ indices) cancel out, leading to
separation of these two equations, (\ref{eqnD24a}) and (\ref{eqnD25a}), into
equations containing only $\omega_{m\left(  pq\right)  }$ and $\omega
_{k\left(  p0\right)  }$, respectively. In addition, when $D=3$, some terms in
the Lagrangian disappear (see \cite{3D}). That is why the Hamiltonian analysis
for $D=3$ becomes so simple, as is the derivation of gauge transformations. In
particular, when $D=3$, equation (\ref{eqnD31}) can be solved because $\left[
\pi^{k\left(  0\right)  }\right]  =\left[  \omega_{m\left(  pq\right)
}\right]  $ $=2$ (see \cite{3D}). Equation (\ref{eqnD31}) for $D=3$ is
represented by two equations for two independent components of $\omega
_{m\left(  pq\right)  }$, $\omega_{1\left(  12\right)  }$ and $\omega
_{2\left(  12\right)  }$. It can be solved for $\omega_{1\left(  12\right)  }$
and $\omega_{2\left(  12\right)  }$ and the solution can be written in
\textquotedblleft covariant\textquotedblright\ form%

\begin{equation}
\omega_{k\left(  qp\right)  }=\frac{1}{2ee^{0\left(  0\right)  }}I_{k\left(
q\right)  m\left(  p\right)  }\pi^{m\left(  0\right)  }. \label{eqnD32}%
\end{equation}

In higher dimensions, equation (\ref{eqnD31}) cannot be solved in the same
way. We showed in \cite{Report} that it is necessary to consider it together
with the secondary constraints. In solving these constraints the combination
of the form $\gamma^{m\left(  n\right)  }\omega_{m\left(  pq\right)  }%
=\tilde{\omega}_{\ \left(  pq\right)  }^{n}$ was very useful (in
$\tilde{\omega}_{\ \left(  pq\right)  }^{n}$ all indices are internal),
because the \textquotedblleft trace\textquotedblright\ of this
combination\footnote{For the original connection $\omega_{m\left(  pq\right)
}$, antisymmetric in internal indices, such a \textquotedblleft
trace\textquotedblright\ cannot be defined. We need combinations with all
indices of the same nature and $\tilde{\omega}_{\ \left(  pq\right)  }^{n}$
(with all indices being internal) provides such a combination which naturally
arises in the direct Hamiltonian analysis \cite{Report}.} is proportional to
$\pi^{k\left(  0\right)  }$%

\begin{equation}
\pi^{k\left(  0\right)  }=2ee^{0\left(  0\right)  }\gamma^{k\left(  p\right)
}\tilde{\omega}_{\ \left(  qp\right)  }^{q}.\label{eqnD33}%
\end{equation}
This suggests the introduction of variables that allow one to single out the
contribution of (\ref{eqnD33}), which is obviously the separation of
$\tilde{\omega}_{n\left(  pq\right)  }$ into the trace, $\tilde{V}_{q}$, and
the traceless, $\tilde{\Omega}_{n\left(  pq\right)  }$, parts%

\begin{equation}
\tilde{\omega}_{n\left(  pq\right)  }=\tilde{\Omega}_{n\left(  pq\right)
}+\frac{1}{D-2}\left(  \tilde{\eta}_{np}\tilde{V}_{q}-\tilde{\eta}_{nq}%
\tilde{V}_{p}\right)  , \label{eqnD34}%
\end{equation}
or equivalently%

\begin{equation}
\omega_{m\left(  pq\right)  }=e_{m}^{\left(  n\right)  }\tilde{\Omega
}_{n\left(  pq\right)  }+\frac{1}{D-2}\left(  e_{m\left(  p\right)  }\tilde
{V}_{q}-e_{m\left(  q\right)  }\tilde{V}_{p}\right)  \label{eqnD35}%
\end{equation}
(where we have used $\tilde{\omega}_{n\left(  pq\right)  }=\tilde{\eta}%
_{nm}\tilde{\omega}_{\ \ \left(  pq\right)  }^{m}$).

The variable $\tilde{\Omega}_{n\left(  pq\right)  }$ is an antisymmetric
($\tilde{\Omega}_{n\left(  pq\right)  }=-\tilde{\Omega}_{n\left(  qp\right)
}$) and traceless ($\tilde{\Omega}_{\ \left(  pq\right)  }^{p}=\tilde{\eta
}^{np}\tilde{\Omega}_{n\left(  pq\right)  }=0$) field with all indices being
internal. The necessary condition for any field redefinition (before checking
the invertability) is that the number of fields is preserved, which is
satisfied in our case because $[\tilde{\omega}_{n\left(  pq\right)  }%
]=[\tilde{\Omega}_{n\left(  pq\right)  }]+\left[  \tilde{V}_{q}\right]  $ in
all dimensions. It is not difficult to demonstrate the invertability of
(\ref{eqnD35}). Contracting (\ref{eqnD34}) with $\gamma^{m\left(  p\right)  }$
(or equally well with $\gamma^{m\left(  q\right)  }$) we obtain%

\begin{equation}
\tilde{V}_{p}=\tilde{\omega}_{\ \left(  qp\right)  }^{q}.\label{eqnD36}%
\end{equation}
Now contracting (\ref{eqnD35}) with $\gamma^{m\left(  k\right)  }$ we find%

\begin{equation}
\tilde{\Omega}_{\ \left(  pq\right)  }^{k}=\gamma^{m\left(  k\right)  }%
\omega_{m\left(  pq\right)  }-\frac{1}{D-2}\left(  \tilde{\delta}_{p}%
^{k}\tilde{V}_{q}-\tilde{\delta}_{q}^{k}\tilde{V}_{p}\right)  \label{eqnD37}%
\end{equation}
and using (\ref{eqnD36})%

\begin{equation}
\tilde{\Omega}_{\ \left(  pq\right)  }^{k}=\gamma^{m\left(  k\right)  }%
\omega_{m\left(  pq\right)  }-\frac{1}{D-2}\left(  \tilde{\delta}_{p}%
^{k}\tilde{\omega}_{\ \left(  nq\right)  }^{n}-\tilde{\delta}_{q}^{k}%
\tilde{\omega}_{\ \left(  np\right)  }^{n}\right)  ,\label{eqnD38}%
\end{equation}
or in terms of the original connections%

\begin{equation}
\tilde{\Omega}_{\ \left(  pq\right)  }^{k}=\gamma^{m\left(  k\right)  }%
\omega_{m\left(  pq\right)  }-\frac{1}{D-2}\left(  \tilde{\delta}_{p}%
^{k}\gamma^{m\left(  n\right)  }\omega_{m\left(  nq\right)  }-\tilde{\delta
}_{q}^{k}\gamma^{m\left(  n\right)  }\omega_{m\left(  np\right)  }\right)  .
\label{eqnD39}%
\end{equation}
We then see that (\ref{eqnD35}) is invertible. It is easy to show also that
(\ref{eqnD39}) is traceless.

We substitute (\ref{eqnD35}) into (\ref{eqnD30}) to express the connection
$\omega_{k\left(  q0\right)  }$ in terms of new fields%

\[
\omega_{k\left(  q0\right)  }=-\frac{1}{2ee^{0\left(  0\right)  }}I_{k\left(
q\right)  m\left(  p\right)  }\pi^{m\left(  p\right)  }-\frac{e^{0\left(
p\right)  }}{2e^{0\left(  0\right)  }}I_{k\left(  q\right)  m\left(  p\right)
}E^{m\left(  a\right)  n\left(  b\right)  }\omega_{n\left(  ab\right)
}\left(  \tilde{\Omega},\tilde{V}\right)  +\frac{e^{0\left(  a\right)  }%
}{e^{0\left(  0\right)  }}\omega_{k\left(  aq\right)  }\left(  \tilde{\Omega
},\tilde{V}\right)
\]
that upon substitution of $\omega_{n\left(  ab\right)  }\left(  \tilde{\Omega
},\tilde{V}\right)  $ and simple contractions gives%

\begin{equation}
\omega_{k\left(  q0\right)  }=-\frac{1}{2ee^{0\left(  0\right)  }}I_{k\left(
q\right)  m\left(  p\right)  }\pi^{m\left(  p\right)  }+\frac{D-3}%
{D-2}e_{k\left(  0\right)  }\tilde{V}_{q}+\frac{e^{0\left(  p\right)  }%
}{e^{0\left(  0\right)  }}e_{k}^{\left(  n\right)  }\tilde{\Omega}_{n\left(
pq\right)  }. \label{eqnD41}%
\end{equation}

Using (\ref{eqnD33}) and (\ref{eqnD36}), we can express $\tilde{V}_{q}$ in
terms of $\pi^{k\left(  0\right)  }$ (this is also linear in auxiliary fields
redefinition with $\left[  \pi^{k\left(  0\right)  }\right]  =\left[
\tilde{V}_{p}\right]  $ in all dimensions)%

\begin{equation}
\tilde{V}_{q}=\frac{1}{2ee^{0\left(  0\right)  }}e_{k\left(  q\right)  }%
\pi^{k\left(  0\right)  }. \label{eqnD42}%
\end{equation}
Finally, we obtain%

\begin{equation}
\omega_{k\left(  q0\right)  }=-\frac{1}{2ee^{0\left(  0\right)  }}I_{k\left(
q\right)  m\left(  p\right)  }\pi^{m\left(  p\right)  }+\frac{D-3}%
{D-2}e_{k\left(  0\right)  }\frac{1}{2ee^{0\left(  0\right)  }}e_{m\left(
q\right)  }\pi^{m\left(  0\right)  }+\frac{e^{0\left(  p\right)  }%
}{e^{0\left(  0\right)  }}e_{k}^{\left(  n\right)  }\tilde{\Omega}_{n\left(
pq\right)  }, \label{eqnD43}%
\end{equation}

\begin{equation}
\omega_{m\left(  pq\right)  }=\frac{1}{D-2}\frac{1}{2ee^{0\left(  0\right)  }%
}\left(  e_{m\left(  p\right)  }e_{n\left(  q\right)  }-e_{m\left(  q\right)
}e_{n\left(  p\right)  }\right)  \pi^{n\left(  0\right)  }+e_{m}^{\left(
n\right)  }\tilde{\Omega}_{n\left(  pq\right)  }. \label{eqnD44}%
\end{equation}
This is a linear transformation (in auxiliary fields) from spatial components
of connections $\omega_{m\left(  \alpha\beta\right)  }$ to the new set of
variables $\pi^{m\left(  \rho\right)  }$and $\tilde{\Omega}_{n\left(
pq\right)  }$.

Note that this field redefinition, (\ref{eqnD43})-(\ref{eqnD44}), equally well
can be performed at the Lagrangian level and \textquotedblleft
momenta\textquotedblright\ $\pi^{m\left(  \rho\right)  }$ are just new
auxiliary variables that play a role of momenta conjugate to $e_{m\left(
\rho\right)  }$ only after passing to the Hamiltonian formulation. In the
Lagrangian formalism, (\ref{eqnD43})-(\ref{eqnD44}) are a definition of
Darboux coordinates, and from now on the auxiliary field $\pi^{m\left(
\rho\right)  }$ will be denoted as $F^{m\left(  \rho\right)  }$.

We can combine (\ref{eqnD43}) and (\ref{eqnD44}) into one \textquotedblleft
semicovariant\textquotedblright\ expression%

\begin{equation}
\omega_{m\left(  \alpha\beta\right)  }=N_{m\left(  \alpha\beta\right)
0n\left(  \sigma\right)  }F^{n\left(  \sigma\right)  }+e_{p\left(
\alpha\right)  }e_{q\left(  \beta\right)  }e_{m}^{\left(  n\right)  }%
\gamma^{p\left(  b\right)  }\gamma^{q\left(  c\right)  }\tilde{\Omega
}_{n\left(  bc\right)  } \label{eqnD45}%
\end{equation}
where%

\begin{equation}
N_{m\left(  \alpha\beta\right)  0n\left(  \sigma\right)  }=\frac
{1}{2ee^{0\left(  0\right)  }}\times\label{eqnD13}%
\end{equation}

\[
\left[  \left(  \tilde{\delta}_{\alpha}^{q}\tilde{\delta}_{\beta}^{0}%
-\tilde{\delta}_{\alpha}^{0}\tilde{\delta}_{\beta}^{q}\right)  \left(
-I_{m\left(  q\right)  n\left(  p\right)  }\tilde{\delta}_{\sigma}^{p}%
+\frac{D-3}{D-2}e_{m\left(  0\right)  }e_{n\left(  q\right)  }\tilde{\delta
}_{\sigma}^{0}\right)  +\tilde{\delta}_{\alpha}^{p}\tilde{\delta}_{\beta}%
^{q}\frac{1}{D-2}\left(  e_{m\left(  p\right)  }e_{n\left(  q\right)
}-e_{m\left(  q\right)  }e_{n\left(  p\right)  }\right)  \tilde{\delta
}_{\sigma}^{0}\right]  .
\]

The advantage of going to Darboux variables%

\begin{equation}
L\left(  e_{\mu\left(  \alpha\right)  },\omega_{\mu\left(  \alpha\beta\right)
}\right)  \rightarrow L\left(  e_{\mu\left(  \alpha\right)  },F^{m\left(
\rho\right)  },\omega_{0\left(  \alpha\beta\right)  },\tilde{\Omega}_{n\left(
pq\right)  }\right)  \label{eqnD46}%
\end{equation}
is based on the following properties%

\begin{equation}
eB^{k\left(  \rho\right)  0\left(  \alpha\right)  m\left(  \beta\right)
}N_{m\left(  \alpha\beta\right)  0n\left(  \sigma\right)  }=\delta_{n}%
^{k}\tilde{\delta}_{\sigma}^{\rho},\text{ \ \ \ }B^{k\left(  \rho\right)
0\left(  \alpha\right)  m\left(  \beta\right)  }\omega_{m\left(  \alpha
\beta\right)  }\left(  \tilde{\Omega}\right)  =0 \label{eqnD47}%
\end{equation}
that for the \textquotedblleft kinetic part\textquotedblright\ of the original
Lagrangian (\ref{eqnD2}) gives a simple expression that is quadratic in fields%

\begin{equation}
eB^{k\left(  \rho\right)  0\left(  \alpha\right)  m\left(  \beta\right)
}e_{k\left(  \rho\right)  ,0}\omega_{m\left(  \alpha\beta\right)
}=F^{k\left(  \rho\right)  }e_{k\left(  \rho\right)  ,0}. \label{eqnD48}%
\end{equation}

The possibility of eliminating the field $\tilde{\Omega}_{c\left(  pq\right)
}$ at the Lagrangian level (Lagrangian reduction) depends on the presence of
terms quadratic in this field. The semicovariant form (\ref{eqnD45}) makes the
calculation quite simple as the only source of a term quadratic in
$\tilde{\Omega}_{c\left(  pq\right)  }$ that contributes in (\ref{eqnD2}) is
the following (see (\ref{eqnD21}))%

\begin{equation}
-eA^{k\left(  \alpha\right)  m\left(  \beta\right)  }\omega_{k\left(
\alpha\gamma\right)  }\omega_{m~\beta)}^{~(\gamma}.\label{eqnD49}%
\end{equation}

Substitution of $\omega_{m\left(  \alpha\beta\right)  }\left(  \tilde{\Omega
}\right)  $ into (\ref{eqnD49}) after contraction with $A^{k\left(
\alpha\right)  m\left(  \beta\right)  }$ gives%

\begin{equation}
L\left(  \tilde{\Omega}\tilde{\Omega}\right)  =e\tilde{\Omega}_{b\left(
pn\right)  }\tilde{\Omega}^{p\left(  nb\right)  }-e\frac{e^{0\left(  a\right)
}}{e^{0\left(  0\right)  }}\tilde{\Omega}_{\ \left(  ap\right)  }^{q}%
\frac{e^{0\left(  b\right)  }}{e_{\left(  0\right)  }^{0}}\tilde{\Omega
}_{\ \left(  bq\right)  }^{p}. \label{eqnD51}%
\end{equation}

Upon performing variation with respect to $\tilde{\Omega}$ an equation similar
to Eq. (R102) follows and, as we demonstrated in \cite{Report}, it can be
solved; though the second term of (\ref{eqnD51}) makes calculations quite long
(note that in the Darboux coordinates (\ref{eqnD45}) we have the equation
(\ref{eqnD51}) as Eq. (R102) immediately, not after long preliminary
calculations as in \cite{Report}). This suggests an additional change of
Darboux coordinates separately for the part proportional to $\tilde{\Omega
}_{b\left(  pn\right)  }$ and the first choice (as we have to keep the number
of components the same) is the antisymmetric traceless field $\hat{\Sigma}%
_{m}^{\ \left(  pq\right)  }$, but with all indices being external and
spatial. Such a field is defined as%

\begin{equation}
\hat{\Sigma}_{m}^{\ \left(  pq\right)  }=e_{m}^{\left(  n\right)  }%
\gamma^{p\left(  b\right)  }\gamma^{q\left(  c\right)  }\tilde{\Omega
}_{n\left(  bc\right)  },\ \ \tilde{\Omega}_{\ \left(  ab\right)  }^{k}%
=\gamma^{m\left(  k\right)  }e_{p\left(  a\right)  }e_{q\left(  b\right)
}\hat{\Sigma}_{m}^{\ \left(  pq\right)  }, \label{eqnD52}%
\end{equation}
which is an invertible redefinition of auxiliary fields.

This additional redefinition diagonalizes (\ref{eqnD51}); and that can be
checked by substitution of (\ref{eqnD52}) into (\ref{eqnD51}), which leads to
only one term that is quadratic in $\hat{\Sigma}_{m}^{\ \left(  pq\right)  }$%

\begin{equation}
L\left(  \tilde{\Omega}\tilde{\Omega}\right)  \Longrightarrow L\left(
\hat{\Sigma}\hat{\Sigma}\right)  =eg_{qp}\hat{\Sigma}_{m}^{\ \left(
kp\right)  }\hat{\Sigma}_{k}^{\ \left(  mq\right)  }\label{eqnD53}%
\end{equation}
(here $g_{qp}$ is a short-hand notation for $e_{q\left(  \alpha\right)  }%
e_{p}^{\left(  \alpha\right)  }$, not an independent field).

This completes the derivation of the Darboux coordinates written down in the
preceding Section in (\ref{eqnD12})%

\begin{equation}
\omega_{m\left(  \alpha\beta\right)  }=N_{m\left(  \alpha\beta\right)
0n\left(  \sigma\right)  }F^{n\left(  \sigma\right)  }+e_{p\left(
\alpha\right)  }e_{q\left(  \beta\right)  }\hat{\Sigma}_{m}^{\ \left(
pq\right)  }=\omega_{m\left(  \alpha\beta\right)  }\left(  F\right)
+\omega_{m\left(  \alpha\beta\right)  }\left(  \hat{\Sigma}\right)  .
\label{eqnD54}%
\end{equation}
The second property of (\ref{eqnD47}) is unaltered by the change of variables
in (\ref{eqnD52}) and%

\begin{equation}
B^{k\left(  \rho\right)  0\left(  \alpha\right)  m\left(  \beta\right)
}\omega_{m\left(  \alpha\beta\right)  }\left(  \hat{\Sigma}\right)  =0.
\label{eqnD55}%
\end{equation}

Using transformation (\ref{eqnD54}) we can now obtain the equivalent
Lagrangian in terms of Darboux coordinates, perform a Lagrangian reduction
(i.e. eliminate $\hat{\Sigma}$) and find the corresponding Hamiltonian as it
was schematically indicated in (\ref{eqnD11}); or equally well, we can start
the Hamiltonian formulation using the Lagrangian in Darboux coordinates and
perform the Hamiltonian reduction. Of course, using either way, we obtain the
same result (\ref{eqnD4})-(\ref{eqnD5}). But before we write the Lagrangian in
Darboux coordinates and the corresponding Hamiltonian we would like to make a
few comments.

In their discussion of Darboux coordinates specialized to the $D=4$ case, the
authors of \cite{Banados} emphasize the non-linearity of their
transformations. Non-linearity in \cite{Banados} and in our (\ref{eqnD12})
appears with respect to only the non-second class fields (the tetrads of
\cite{Banados} or N-beins in our case), and exactly opposite, linearity in the
second class fields ($\pi^{k\left(  \rho\right)  },\hat{\lambda}_{km}$ in
\cite{Banados} and $F^{n\left(  \sigma\right)  },\hat{\Sigma}_{m}^{\ \left(
pq\right)  }$ in our case), making the transformation invertible which is a
necessary condition to establish equivalence of the original formulation with
the formulation in terms of Darboux coordinates.

In constructing of (\ref{eqnD13}) we used the results of the Hamiltonian
analysis that preserves the $D=3$ limit. A different, \textquotedblleft more
covariant\textquotedblright, combination can be constructed that also
diagonalizes the \textquotedblleft kinetic\textquotedblright\ part of the
Lagrangian (i.e. has the same properties as (\ref{eqnD13})); for example%

\begin{equation}
N_{m\left(  \alpha\beta\right)  0n\left(  \sigma\right)  }^{\prime}=\frac
{1}{e}\left[  e_{m\left(  \sigma\right)  }A_{n\left(  \alpha\right)  0\left(
\beta\right)  }-\frac{1}{D-2}\left(  e_{m\left(  \alpha\right)  }A_{n\left(
\sigma\right)  0\left(  \beta\right)  }-e_{m\left(  \beta\right)  }A_{n\left(
\sigma\right)  0\left(  \alpha\right)  }\right)  \right]  \label{eqnD56}%
\end{equation}
where%

\begin{equation}
A_{n\left(  \alpha\right)  0\left(  \beta\right)  }=e_{n\left(  \alpha\right)
}e_{0\left(  \beta\right)  }-e_{n\left(  \beta\right)  }e_{0\left(
\alpha\right)  }. \label{eqnD57}%
\end{equation}

Without any preliminary Hamiltonian analysis, and working only in a particular
dimension (e.g. $D=4$), such a diagonalization of the \textquotedblleft
kinetic\textquotedblright\ part looks even preferable as it has a
\textquotedblleft more covariant\textquotedblright\ form that simplifies
calculations and does not involve a division by $e^{0\left(  0\right)  }$ as
in (\ref{eqnD13}). However, it does not have the correct $D=3$ limit, which
cannot be seen if one is working only in a particular dimension, e.g. when
$D=4$ (see point C in the Introduction). Using the Darboux coordinates
(\ref{eqnD54}) with $N_{m\left(  \alpha\beta\right)  0n\left(  \sigma\right)
}^{\prime}$, instead of $N_{m\left(  \alpha\beta\right)  0n\left(
\sigma\right)  }$, leads to problems in the Hamiltonian analysis. The reason
for this is that the transformation found using the Hamiltonian analysis,
(\ref{eqnD13}), preserves properties of the primary variables, i.e.
$\frac{\delta\omega_{m\left(  \alpha\beta\right)  }}{\delta e_{0\left(
\rho\right)  }}=0$, for the Darboux coordinates with $N_{m\left(  \alpha
\beta\right)  0n\left(  \sigma\right)  }$, as%
\begin{equation}
\frac{\delta}{\delta e_{0\left(  \rho\right)  }}\left(  N_{m\left(
\alpha\beta\right)  0n\left(  \sigma\right)  }\pi^{n\left(  \sigma\right)
}+e_{p\left(  \alpha\right)  }e_{q\left(  \beta\right)  }\hat{\Sigma}%
_{m}^{\ \left(  pq\right)  }\right)  =0,\label{eqnD58}%
\end{equation}
in contrast to the \textquotedblleft more covariant\textquotedblright%
\ combination $N_{m\left(  \alpha\beta\right)  0n\left(  \sigma\right)
}^{\prime}$ for which%

\begin{equation}
\frac{\delta}{\delta e_{0\left(  \rho\right)  }}N_{m\left(  \alpha
\beta\right)  0n\left(  \sigma\right)  }^{\prime}\neq0.\label{eqnD59}%
\end{equation}
This leads to change of the algebra of constraints, gauge invariance, etc.

This is the illustration of how a pure Lagrangian consideration of singular
systems can destroy its properties if one assumes that one can always use some
operations (e.g. any invertible transformation) known for non-singular
Lagrangians without careful analysis and without taking into account the
specifics of singular systems (see a general discussion in \cite{GGT}). When
constructing the Darboux coordinates, we rely on the Hamiltonian analysis\ and
this protects us from such mistakes.\footnote{This example illustrates why the
problems might arise in the Faddeev-Jackiw method \cite{FJ}. The simplectic
form in \cite{FJ} was found by diagonalizing the \textquotedblleft
kinetic\textquotedblright\ part of the Lagrangian but, as we have shown, it is
not enough in general, to preserve equivalence and this is the reason for
\textquotedblleft non-equivalence\textquotedblright\ of the Dirac and
symplectic methods found for some models (see e.g. \cite{RR}, \cite{Shirzad}
where role of second class constraints was emphasized, and \cite{Pons} where
the observation of non-equivalence leads the authors to the conclusion about
deficiency of the Dirac method).} Construction of Darboux coordinates for one
particular theory, the first order EC action, is the illustration of Dirac's
general statement \cite{Diracbook}: \textquotedblleft I [Dirac] feel that
there will always be something missing from them [non-Hamiltonian methods]
which we can only get by working from a Hamiltonian\textquotedblright.\ 

\section{Lagrangian and Hamiltonian reductions of the EC theory in Darboux
coordinates}

Substitution of the Darboux coordinates (\ref{eqnD54}) into the original EC
Lagrangian $L_{EC}$ (\ref{eqnD2}) is a simple task, as we have only one
expression for all spatial connections (\ref{eqnD54}) which are the only
fields that are affected by a change of variables. This gives us a different
but equivalent first order formulation of the EC theory $L_{EC\left(
D\right)  }$%

\[
L_{EC}\left(  e_{\mu\left(  \alpha\right)  },\omega_{\mu\left(  \alpha
\beta\right)  }\right)  \Longrightarrow L_{EC\left(  D\right)  }\left(
e_{\mu\left(  \alpha\right)  },\omega_{0\left(  \alpha\beta\right)
},F^{k\left(  \rho\right)  },\hat{\Sigma}_{m}^{\ \left(  pq\right)  }\right)
=
\]

\[
e_{k\left(  \rho\right)  ,0}F^{k\left(  \rho\right)  }+\left(  eB^{k\left(
\rho\right)  m\left(  \alpha\right)  0\left(  \beta\right)  }e_{k\left(
\rho\right)  ,m}-2eA^{0\left(  \alpha\right)  k\left(  \gamma\right)  }%
\omega_{k~~\gamma)}^{~(\beta}\left(  F,\hat{\Sigma}\right)  \right)
\omega_{0\left(  \alpha\beta\right)  }%
\]

\begin{equation}
-e_{0\left(  \rho\right)  ,k}F^{k\left(  \rho\right)  }+eB^{n\left(
\rho\right)  k\left(  \alpha\right)  m\left(  \beta\right)  }e_{n\left(
\rho\right)  ,k}\omega_{m\left(  \alpha\beta\right)  }\left(  F,\hat{\Sigma
}\right)  -eA^{k\left(  \alpha\right)  m\left(  \beta\right)  }\omega
_{k\left(  \alpha\gamma\right)  }\left(  F,\hat{\Sigma}\right)  \omega
_{m~~\beta)}^{~(\gamma}\left(  F,\hat{\Sigma}\right)  . \label{eqnD60}%
\end{equation}

The appearance of two terms that are quadratic in the fields (first terms in
the second and third lines of (\ref{eqnD60})) is the consequence of
(\ref{eqnD48}). Further separating spatial connections into two parts (see
(\ref{eqnD43}), (\ref{eqnD44})) and performing some contractions we obtain%

\[
L_{EC\left(  D\right)  }=e_{k\left(  \rho\right)  ,0}F^{k\left(  \rho\right)
}+\left(  \frac{1}{2}F^{k\left(  \alpha\right)  }e_{k}^{\left(  \beta\right)
}-\frac{1}{2}F^{k\left(  \beta\right)  }e_{k}^{\left(  \alpha\right)
}+eB^{k\left(  \rho\right)  m\left(  \alpha\right)  0\left(  \beta\right)
}e_{k\left(  \rho\right)  ,m}\right)  \omega_{0\left(  \alpha\beta\right)  }%
\]

\begin{equation}
-e_{0\left(  \rho\right)  ,k}F^{k\left(  \rho\right)  }+eB^{n\left(
\rho\right)  k\left(  \alpha\right)  m\left(  \beta\right)  }e_{n\left(
\rho\right)  ,k}\omega_{m\left(  \alpha\beta\right)  }\left(  F\right)
-eA^{k\left(  \alpha\right)  m\left(  \beta\right)  }\omega_{k\left(
\alpha\gamma\right)  }\left(  F\right)  \omega_{m~~\beta)}^{~(\gamma}\left(
F\right)  \label{eqnD61}%
\end{equation}

\[
+eg_{qp}\hat{\Sigma}_{m}^{\ \left(  kp\right)  }\hat{\Sigma}_{k}^{\ \left(
mq\right)  }+2ee^{k\left(  \beta\right)  }\left(  e_{m\left(  \beta\right)
,q}+e_{q}^{\left(  \gamma\right)  }\omega_{m\left(  \gamma\beta\right)
}\left(  F\right)  \right)  \hat{\Sigma}_{k}^{\ \left(  mq\right)  },
\]
where $g_{qp}$ is, as before, a short-hand notation for $e_{q\left(
\rho\right)  }e_{p}^{\left(  \rho\right)  },$ not an independent field. Note
that in the terms proportional to $\omega_{0\left(  \alpha\beta\right)  }$,
there are no contributions involving $\hat{\Sigma}_{k}^{\ \left(  mq\right)
}$ as $A^{0\left(  \alpha\right)  k\left(  \gamma\right)  }\omega_{k~~\gamma
)}^{~(\beta}\left(  \hat{\Sigma}\right)  =0$; and contributions with
$F^{k\left(  \alpha\right)  }$, instead of direct substitution, can be
obtained by contracting $B^{k\left(  \rho\right)  m\left(  \alpha\right)
0\left(  \beta\right)  }$ with $e_{k}^{\left(  \beta\right)  }$ and performing
an antisymmetrization that gives $eA^{0\left(  \alpha\right)  k\left(
\gamma\right)  }\omega_{k~~\gamma)}^{~(\beta}\left(  F\right)  $. The last
line of (\ref{eqnD61}) is the result of a contraction with the explicit form
of $\omega_{k\left(  \alpha\gamma\right)  }\left(  \hat{\Sigma}\right)  $ (see
(\ref{eqnD54})).

The last line of (\ref{eqnD61}) is the algebraic expression with respect to
the field $\hat{\Sigma}_{k}^{\ \left(  mq\right)  }$ that can be eliminated
(due to the presence of a term in the Lagrangian quadratic in this field) by
using its equation of motion (Lagrangian reduction). After elimination of this
field we can obtain the Lagrangian (see (\ref{eqnD9})) with $F^{k\left(
\rho\right)  }$ playing the role of momenta conjugate to $e_{k\left(
\rho\right)  }$ in the Hamiltonian formulation, and without the need to solve
the secondary second class constraints. Thus, using the Darboux coordinates we
can obtain (\ref{eqnD4})-(\ref{eqnD5}), which is the same Hamiltonian derived
in \cite{Report}, but without having to do any long calculation. Equivalence
of the Lagrangian and Hamiltonian methods allows us to interchange the order
of operations, and using (\ref{eqnD61}) (without the Lagrangian reduction) we
can immediately write the Hamiltonian and then perform the Hamiltonian
reduction. This approach is employed in the present paper, giving us the
possibility of comparing the results obtained here with the direct
calculations of \cite{Report}.

The Hamiltonian of the first order EC action written in Darboux coordinates
can be just read off from the corresponding Lagrangian (\ref{eqnD61}), as is
possible for any first order action. The advantage of Darboux coordinates is
that the primary constraints are very simple and the second class variables
$F^{k\left(  \alpha\right)  }$ and $\hat{\Sigma}_{k}^{\ \left(  mq\right)  }$
can be easily separated and eliminated by solving the second class constraints
(Hamiltonian reduction).

The total Hamiltonian (introducing momenta conjugate to all fields in the
Lagrangian) is%

\[
H_{T}\left(  e_{\mu\left(  \rho\right)  },\pi^{\mu\left(  \rho\right)
},\omega_{0\left(  \alpha\beta\right)  },\Pi^{0\left(  \alpha\beta\right)
},F^{k\left(  \alpha\right)  },\Pi_{k\left(  \alpha\right)  },\hat{\Sigma}%
_{k}^{\ \left(  mq\right)  },\hat{\Pi}_{\ \ \left(  mq\right)  }^{k}\right)
\]

\begin{equation}
=\pi^{\mu\left(  \rho\right)  }\dot{e}_{\mu\left(  \rho\right)  }%
+\Pi^{0\left(  \alpha\beta\right)  }\dot{\omega}_{0\left(  \alpha\beta\right)
}+\dot{F}^{k\left(  \alpha\right)  }\Pi_{k\left(  \alpha\right)  }+\hat
{\Sigma}_{k,0}^{\ \left(  mq\right)  }\hat{\Pi}_{\ \ \left(  mq\right)  }%
^{k}-L~,\label{eqnD63}%
\end{equation}
where $\pi^{\mu\left(  \rho\right)  }$, $\Pi^{0\left(  \alpha\beta\right)  }$,
$\Pi_{k\left(  \alpha\right)  }$ and $\hat{\Pi}_{\ \ \left(  mq\right)  }^{k}$
are momenta conjugate to $e_{\mu\left(  \rho\right)  },$ $\omega_{0\left(
\alpha\beta\right)  }$, $F^{k\left(  \alpha\right)  }$ and $\hat{\Sigma}%
_{k}^{\ \left(  mq\right)  }$, respectively. Separating terms with
\textquotedblleft velocities\textquotedblright\ in the Lagrangian
(\ref{eqnD61}) we write%

\[
-L=-e_{k\left(  \rho\right)  ,0}F^{k\left(  \rho\right)  }+H_{c}%
\]
and then singling out terms proportional to $\omega_{0\left(  \alpha
\beta\right)  }$ we obtain%

\begin{equation}
H_{c}=-\omega_{0\left(  \alpha\beta\right)  }\chi^{0\left(  \alpha
\beta\right)  }+H_{c}^{^{\prime}} \label{eqnD64}%
\end{equation}
where%

\begin{equation}
\chi^{0\left(  \alpha\beta\right)  }=\frac{1}{2}F^{k\left(  \alpha\right)
}e_{k}^{\left(  \beta\right)  }-\frac{1}{2}F^{k\left(  \beta\right)  }%
e_{k}^{\left(  \alpha\right)  }+eB^{k\left(  \rho\right)  m\left(
\alpha\right)  0\left(  \beta\right)  }e_{k\left(  \rho\right)  ,m}
\label{eqnD65}%
\end{equation}
and%

\[
H_{c}^{^{\prime}}=e_{0\left(  \rho\right)  ,k}F^{k\left(  \rho\right)
}-eB^{n\left(  \rho\right)  k\left(  \alpha\right)  m\left(  \beta\right)
}e_{n\left(  \rho\right)  ,k}\omega_{m\left(  \alpha\beta\right)  }\left(
F\right)  +eA^{k\left(  \alpha\right)  m\left(  \beta\right)  }\omega
_{k\left(  \alpha\gamma\right)  }\left(  F\right)  \omega_{m~\beta)}%
^{~(\gamma}\left(  F\right)
\]

\begin{equation}
-eg_{qp}\hat{\Sigma}_{k}^{\ \left(  mq\right)  }\hat{\Sigma}_{m}^{\ \left(
kp\right)  }-2ee^{k\left(  \beta\right)  }\left(  e_{m\left(  \beta\right)
,q}+e_{q}^{\left(  \gamma\right)  }\omega_{m\left(  \gamma\beta\right)
}\left(  F\right)  \right)  \hat{\Sigma}_{k}^{\ \left(  mq\right)  }.
\label{eqnD66}%
\end{equation}

The non-zero fundamental PBs are%

\begin{equation}
\left\{  e_{\nu\left(  \sigma\right)  },\pi^{\mu\left(  \rho\right)
}\right\}  =\delta_{\nu}^{\mu}\tilde{\delta}_{\sigma}^{\rho},\text{
\ \ }\left\{  \omega_{0\left(  \alpha\beta\right)  },\Pi^{0\left(  \rho
\sigma\right)  }\right\}  =\tilde{\Delta}_{\left(  \alpha\beta\right)
}^{\left(  \rho\sigma\right)  },\text{ \ }\left\{  F^{k\left(  \alpha\right)
},\Pi_{m\left(  \beta\right)  }\right\}  =\delta_{m}^{k}\tilde{\delta}_{\beta
}^{\alpha}, \label{eqnD67}%
\end{equation}

\begin{equation}
\left\{  \hat{\Sigma}_{k}^{\ \left(  mq\right)  },\hat{\Pi}_{\ \left(
yz\right)  }^{x}\right\}  =\hat{I}_{k\left(  yz\right)  }^{x\left(  mq\right)
}=\delta_{k}^{x}\hat{\Delta}_{\left(  yz\right)  }^{\left(  mq\right)  }%
-\frac{1}{D-2}\left(  \delta_{y}^{x}\hat{\Delta}_{\left(  kz\right)
}^{\left(  mq\right)  }-\delta_{z}^{x}\hat{\Delta}_{\left(  ky\right)
}^{\left(  mq\right)  }\right)  \label{eqnD68}%
\end{equation}
where%

\begin{equation}
\tilde{\Delta}_{\left(  \alpha\beta\right)  }^{\left(  \rho\sigma\right)
}\equiv\frac{1}{2}\left(  \tilde{\delta}_{\alpha}^{\rho}\tilde{\delta}_{\beta
}^{\sigma}-\tilde{\delta}_{\beta}^{\rho}\tilde{\delta}_{\alpha}^{\sigma
}\right)  ,\text{ \ \ }\hat{\Delta}_{\left(  yz\right)  }^{\left(  mq\right)
}\equiv\frac{1}{2}\left(  \delta_{y}^{m}\delta_{z}^{q}-\delta_{z}^{m}%
\delta_{y}^{q}\right)  .\label{eqnD69}%
\end{equation}

As in any first order formulation, the number of primary constraints is equal
to the number of independent variables. One pair of primary constraints%

\begin{equation}
\phi^{k\left(  \rho\right)  }=\pi^{k\left(  \rho\right)  }-F^{k\left(
\rho\right)  }\approx0\text{, \ }\Pi_{m\left(  \gamma\right)  }\approx
0\text{\ } \label{eqn69a}%
\end{equation}
is second class. These are constraints of a special form \cite{GT} and one
pair of phase-space variables can be eliminated without affecting the PBs of
the remaining variables by substitution of the solution into the total Hamiltonian%

\begin{equation}
F^{k\left(  \rho\right)  }=\pi^{k\left(  \rho\right)  }\text{, \ }%
\Pi_{m\left(  \gamma\right)  }=0. \label{eqn69b}%
\end{equation}

This is the first stage of Hamiltonian reduction and illustrates the
classification (suggested in the previous Section) on primary and second class
variables: $F^{k\left(  \rho\right)  }$ is a second class variable. After this
reduction the total Hamiltonian is\ %

\begin{equation}
H_{T}=\pi^{0\left(  \rho\right)  }\dot{e}_{0\left(  \rho\right)  }%
+\Pi^{0\left(  \alpha\beta\right)  }\dot{\omega}_{0\left(  \alpha\beta\right)
}+\hat{\Sigma}_{k,0}^{\ \left(  mq\right)  }\hat{\Pi}_{\ \ \left(  mq\right)
}^{k}-\omega_{0\left(  \alpha\beta\right)  }\chi^{0\left(  \alpha\beta\right)
}\left(  F=\pi\right)  +H_{c}^{^{\prime}}\left(  F=\pi\right)  .
\label{eqnD69c}%
\end{equation}

According to the Dirac procedure, the next step is to consider the time
development of the primary constraints ($\Pi^{0\left(  \alpha\beta\right)  }$,
$\pi^{0\left(  \rho\right)  }$, $\hat{\Pi}_{\ \ \left(  mq\right)  }^{k}$).
After the first reduction, all primary constraints obviously have zero PBs
among themselves (they are momenta of canonical variables), i.e. there are no
second class pairs among the primary constraints, and all of them lead to the
corresponding secondary constraints, e.g.%

\begin{equation}
\dot{\Pi}^{0\left(  \alpha\beta\right)  }=\left\{  \Pi^{0\left(  \alpha
\beta\right)  },H_{T}\right\}  =\left\{  \Pi^{0\left(  \alpha\beta\right)
},H_{c}\right\}  =\left\{  \Pi^{0\left(  \alpha\beta\right)  },-\omega
_{0\left(  \alpha\beta\right)  }\chi^{0\left(  \alpha\beta\right)  }\right\}
=\chi^{0\left(  \alpha\beta\right)  }. \label{eqnD70}%
\end{equation}
The secondary rotational constraint $\chi^{0\left(  \alpha\beta\right)  }$ has
zero PBs with all primary constraints%

\begin{equation}
\left\{  \chi^{0\left(  \alpha\beta\right)  },\Pi^{0\left(  \nu\mu\right)
}\right\}  =\left\{  \chi^{0\left(  \alpha\beta\right)  },\pi^{0\left(
\sigma\right)  }\right\}  =\left\{  \chi^{0\left(  \alpha\beta\right)  }%
,\hat{\Pi}_{\ \ \left(  mq\right)  }^{k}\right\}  =0.\label{eqnD71}%
\end{equation}
The first and last PBs are manifestly zero and the second is just the
consequence of the properties of $B^{\lambda\left(  \gamma\right)  \mu\left(
\alpha\right)  \nu\left(  \beta\right)  }$ and antisymmetry of $C^{\tau\left(
\sigma\right)  \lambda\left(  \gamma\right)  \mu\left(  \alpha\right)
\nu\left(  \beta\right)  }$ (see (\ref{eqn41a}) of Appendix A)%

\begin{equation}
\frac{\delta}{\delta e_{0\left(  \sigma\right)  }}\left(  eB^{k\left(
\rho\right)  m\left(  \alpha\right)  0\left(  \beta\right)  }e_{k\left(
\rho\right)  ,m}\right)  =eC^{0\left(  \sigma\right)  k\left(  \rho\right)
m\left(  \alpha\right)  0\left(  \beta\right)  }e_{k\left(  \rho\right)
,m}=0. \label{eqnD72}%
\end{equation}
The PB among two rotational constraints coincides with Lorentz algebra
(\ref{eqnD6}); this has been already demonstrated in \cite{3D, Report} for all dimensions.

Because $\left\{  \pi^{0\left(  \rho\right)  },\chi^{0\left(  \alpha
\beta\right)  }\right\}  =0$ and since the PBs among all primary constraint
(\ref{eqnD70}) all vanish, the time development of $\pi^{0\left(
\sigma\right)  }$ leads to the secondary translational constraint%

\begin{equation}
\dot{\pi}^{0\left(  \sigma\right)  }=\left\{  \pi^{0\left(  \sigma\right)
},H_{T}\right\}  =\left\{  \pi^{0\left(  \sigma\right)  },H_{c}\right\}
=\left\{  \pi^{0\left(  \sigma\right)  },H_{c}^{^{\prime}}\right\}
=-\frac{\delta H_{c}^{^{\prime}}}{\delta e_{0\left(  \sigma\right)  }}%
=\chi^{0\left(  \sigma\right)  }, \label{eqnD73}%
\end{equation}
which has the explicit form%

\[
\chi^{0\left(  \sigma\right)  }=\pi_{,k}^{k\left(  \sigma\right)
}+eC^{0\left(  \sigma\right)  n\left(  \rho\right)  k\left(  \alpha\right)
m\left(  \beta\right)  }e_{n\left(  \rho\right)  ,k}\omega_{m\left(
\alpha\beta\right)  }\left(  \pi\right)  -eB^{0\left(  \sigma\right)  k\left(
\alpha\right)  m\left(  \beta\right)  }\omega_{k\left(  \alpha\gamma\right)
}\left(  \pi\right)  \omega_{m~\beta)}^{~(\gamma}\left(  \pi\right)
\]

\begin{equation}
+ee^{0\left(  \sigma\right)  }g_{qp}\hat{\Sigma}_{k}^{\ \left(  mq\right)
}\hat{\Sigma}_{m}^{\ \left(  kp\right)  }+2eA^{0\left(  \sigma\right)
k\left(  \beta\right)  }\left(  e_{m\left(  \beta\right)  ,q}+e_{q}^{\left(
\gamma\right)  }\omega_{m\left(  \gamma\beta\right)  }\left(  \pi\right)
\right)  \hat{\Sigma}_{k}^{\ \left(  mq\right)  }. \label{eqnD74}%
\end{equation}

When performing the variation in (\ref{eqnD73}) we used $\frac{\delta
\omega_{k\left(  \gamma\beta\right)  }\left(  \pi\right)  }{\delta e_{0\left(
\sigma\right)  }}=0$ (this is easy to show using $\omega_{k\left(  \gamma
\beta\right)  }\left(  \pi\right)  $ from (\ref{eqnD54}) and the fact that
$\frac{\delta}{\delta e_{0\left(  \sigma\right)  }}\left(  \frac
{1}{ee^{0\left(  0\right)  }}\right)  =0$ and $\frac{\delta e_{k\left(
\mu\right)  }}{\delta e_{0\left(  \sigma\right)  }}=0$). In all terms in
$H_{c}^{^{\prime}}$, only $ABC$ densities are affected and their variations
are simple (see Appendix A).

Contracting (\ref{eqnD74}) with $e_{0\left(  \sigma\right)  }$ and using the
$ABC$ properties (expand $A$ and $B$ in $\sigma$ and contract with
$e_{0\left(  \sigma\right)  }$ (see (\ref{eqn42a}) of Appendix A)), we can
express $H_{c}^{\prime}$ as
\begin{equation}
H_{c}^{\prime}=-e_{0\left(  \sigma\right)  }\chi^{0\left(  \sigma\right)
}+\left(  e_{0\left(  \rho\right)  }\pi^{k\left(  \rho\right)  }\right)
_{,k}. \label{eqnD75}%
\end{equation}

Based on the properties of the $ABC$ functions we immediately obtain that the
PB of $\chi^{0\left(  \sigma\right)  }$ with the primary translational
constraint, $\pi^{0\left(  \mu\right)  }$, is zero (i.e. after second
variation with respect to $e_{0\left(  \mu\right)  }$ we will have $A,B,C$ and
$D$ with two equal indices $(00)$ which are zero because of antisymmetry of
these density functions):%

\begin{equation}
\left\{  \chi^{0\left(  \sigma\right)  },\pi^{0\left(  \mu\right)  }\right\}
=0, \label{eqnD76}%
\end{equation}
and also%

\begin{equation}
\left\{  \chi^{0\left(  \sigma\right)  },\Pi^{0\left(  \alpha\beta\right)
}\right\}  =0. \label{eqnD77}%
\end{equation}

The PB of $\chi^{0\left(  \sigma\right)  }$ with the primary constraint
$\hat{\Pi}_{\ \ \left(  mq\right)  }^{k}$ is not zero, but the time
development of $\hat{\Pi}_{\ \ \left(  mq\right)  }^{k}$ leads to the
secondary constraint%

\begin{equation}
\hat{\Pi}_{\ \ \left(  mq\right)  ,0}^{k}=\left\{  \hat{\Pi}_{\ \ \left(
mq\right)  }^{k},H_{c}^{^{\prime}}\right\}  =-\frac{\delta H_{c}^{^{\prime}}%
}{\delta\hat{\Sigma}_{k}^{\ \left(  mq\right)  }}=\hat{\chi}_{\ \left(
mq\right)  }^{k}\label{eqnD78}%
\end{equation}
where%

\begin{equation}
\hat{\chi}_{\ \left(  mq\right)  }^{k}=e\hat{\Sigma}_{m}^{\ \left(  kp\right)
}g_{pq}-e\hat{\Sigma}_{q}^{\ \left(  kp\right)  }g_{pm}+e\hat{D}_{\ \left(
mq\right)  }^{k}\label{eqnD79}%
\end{equation}
with $\hat{D}_{\ \left(  mq\right)  }^{k}$ being the manifestly antisymmetric
and traceless combination (see Appendix B)%

\begin{equation}
\hat{D}_{\ \left(  mq\right)  }^{k}=\hat{D}_{\ mq}^{k}-\hat{D}_{\ qm}%
^{k}-\frac{1}{D-2}\left[  \delta_{m}^{k}\left(  \hat{D}_{\ nq}^{n}-\hat
{D}_{\ qn}^{n}\right)  -\delta_{q}^{k}\left(  \hat{D}_{\ nm}^{n}-\hat
{D}_{\ mn}^{n}\right)  \right]  , \label{eqnD80}%
\end{equation}
built from the coefficient which appears in front of terms in (\ref{eqnD66})
that are linear in $\hat{\Sigma}_{k}^{\ \left(  mq\right)  }$%

\begin{equation}
\hat{D}_{\ mq}^{k}=e^{k\left(  \beta\right)  }e_{m\left(  \beta\right)
,q}+e^{k\left(  \beta\right)  }e_{q}^{\left(  \gamma\right)  }\omega_{m\left(
\gamma\beta\right)  }\left(  \pi\right)  . \label{eqnD81}%
\end{equation}

Note that $\hat{D}_{\ mq}^{k}$ by itself is not antisymmetric or traceless;
and (\ref{eqnD79})-(\ref{eqnD80}) are the result of performing variation using
the fundamental PB of (\ref{eqnD68}).

The pair of constraints ($\hat{\Pi}_{\ \ \left(  mq\right)  }^{k},\hat{\chi
}_{\ \ \left(  yz\right)  }^{x}$) is second class because%

\begin{equation}
\left\{  \hat{\Pi}_{\ \ \left(  mq\right)  }^{k},\hat{\chi}_{\ \ \left(
yz\right)  }^{x}\right\}  =\hat{N}_{\left(  mq\right)  \left(  yz\right)
}^{kx}, \label{eqnD82}%
\end{equation}
where%

\begin{equation}
\hat{N}_{\left(  mq\right)  \left(  yz\right)  }^{kx}=-\frac{e}{2}\left[
\delta_{y}^{k}\left(  g_{zm}\delta_{q}^{x}-g_{zq}\delta_{m}^{x}\right)
-\delta_{z}^{k}\left(  g_{ym}\delta_{q}^{x}-g_{yq}\delta_{m}^{x}\right)
\right.  \label{eqnD83}%
\end{equation}

\[
\left.  -\frac{1}{D-2}\left[  \delta_{y}^{x}\left(  g_{zm}\delta_{q}%
^{k}-g_{zq}\delta_{m}^{k}\right)  -\delta_{z}^{x}\left(  g_{ym}\delta_{q}%
^{k}-g_{yq}\delta_{m}^{k}\right)  \right]  \right]  ,
\]
which is manifestly antisymmetric in $\left(  mq\right)  $ and $\left(
yz\right)  $, traceless in $km$ and $kq,$ in $xy$ and $xz$, non-zero and not
proportional to constraints. (Here $g_{zm}$ again denotes $e_{z\left(
\alpha\right)  }e_{m}^{\left(  \alpha\right)  }$.) The most important property
of $\hat{N}_{\left(  mq\right)  \left(  yz\right)  }^{kx}$ is its
invertability with the explicit form of inverse is given below. The pair of
variables ($\hat{\Pi}_{\ \ \left(  mq\right)  }^{k},\hat{\Sigma}%
_{k}^{\ \left(  mq\right)  }$) can be eliminated by substitution into the
total Hamiltonian%

\begin{equation}
\hat{\Pi}_{\ \ \left(  mq\right)  }^{k}=0,\text{ \ \ }\hat{\Sigma}%
_{z}^{\ \left(  mk\right)  }=\frac{1}{2}\left(  \gamma^{kn}\hat{D}_{\ \left(
nz\right)  }^{m}-\gamma^{mn}\hat{D}_{\ \left(  nz\right)  }^{k}-\gamma
^{ky}\gamma^{mw}g_{zx}\hat{D}_{\ \left(  wy\right)  }^{x}\right)
,\label{eqnD84}%
\end{equation}
where $\hat{\Sigma}_{z}^{\ \left(  mk\right)  }$ is the solution of the
constraint (\ref{eqnD79}) $\hat{\chi}_{\ \left(  mq\right)  }^{k}=0$ (see
Appendix B). This again illustrates our classification: $\hat{\Sigma}%
_{k}^{\ \left(  mq\right)  }$ is a second class variable as is $F^{k\left(
\rho\right)  }$. In (\ref{eqnD84}) we use a short-hand notation which was
originally introduced by Dirac \cite{Dirac} for the Hamiltonian formulation of
the Einstein-Hilbert action: $\gamma^{kn}\equiv g^{kn}-\frac{g^{0k}g^{0n}%
}{g^{00}}$ where $g^{\mu\nu}=e^{\mu\left(  \alpha\right)  }e_{\left(
\alpha\right)  }^{\nu}$.

The elimination of the phase-space pair ($F^{k\left(  \rho\right)  }$,
$\Pi_{k\left(  \rho\right)  }$) by solving the corresponding second class
constraint was simple as they are of a special form and it is known that in
such a case the Dirac brackets (DBs) of the remaining fields coincide with
their original PBs \cite{GT}. The pair of second class constraints ($\hat{\Pi
}_{\ \ \left(  mq\right)  }^{k},\hat{\chi}_{\ \ \left(  yz\right)  }^{x}$) is
more complicated and the effect of their elimination on the PBs among the
remaining canonical variables has to be checked. In \cite{Report} the
elimination of $\omega_{k\left(  \alpha\beta\right)  }$ was performed using a
different and complicated approach because of the mixture of different
components ($\omega_{k\left(  pq\right)  }$ and $\omega_{k\left(  p0\right)
}$) in the equations, and it was even not clear how DBs can be calculated.
After introducing Darboux coordinates and decoupling the two fields,
$F^{k\left(  \rho\right)  }$ and $\hat{\Sigma}_{k}^{\ \left(  mq\right)  }$,
this is possible. Let us investigate the effect of their elimination on the
DBs of the remaining fields.

The Dirac bracket is defined for any pair of functions of canonical variables
as \cite{Diracbook}%

\begin{equation}
\left\{  \Phi,\Psi\right\}  _{DB}=\left\{  \Phi,\Psi\right\}  _{PB}-\left(
\begin{array}
[c]{cc}%
\left\{  \Phi,\hat{\Pi}_{~\left(  mq\right)  }^{k}\right\}  _{PB} & \left\{
\Phi,\hat{\chi}_{~\left(  mq\right)  }^{k}\right\}  _{PB}%
\end{array}
\right)  M^{-1}\left(
\begin{array}
[c]{c}%
\left\{  \hat{\Pi}_{~\left(  bc\right)  }^{a},\Psi\right\}  _{PB}\\
\left\{  \hat{\chi}_{~\left(  bc\right)  }^{a},\Psi\right\}  _{PB}%
\end{array}
\right)  .\label{eqnD85}%
\end{equation}
(Here we used our set of second class constraints.) $M^{-1}$ is the inverse of
the matrix $M$ built from the PBs of the second class constraints%

\begin{equation}
M=\left[
\begin{array}
[c]{cc}%
\left\{  \hat{\Pi}_{~\left(  mq\right)  }^{k},\hat{\Pi}_{~\left(  bc\right)
}^{a}\right\}  & \left\{  \hat{\Pi}_{\text{~}\left(  mq\right)  }^{k}%
,\hat{\chi}_{~\left(  bc\right)  }^{a}\right\} \\
\left\{  \hat{\chi}_{~\left(  mq\right)  }^{k},\hat{\Pi}_{~\left(  bc\right)
}^{a}\right\}  & \left\{  \hat{\chi}_{~\left(  mq\right)  }^{k},\hat{\chi
}_{~\left(  bc\right)  }^{a}\right\}
\end{array}
\right]  =\left[
\begin{array}
[c]{cc}%
0 & \hat{N}_{\left(  mq\right)  \left(  bc\right)  }^{ka}\\
-\hat{N}_{\left(  bc\right)  \left(  mq\right)  }^{ak} & \hat{X}_{\left(
mq\right)  \left(  bc\right)  }^{ka}%
\end{array}
\right]  . \label{eqnD86}%
\end{equation}
Note that $\hat{N}_{\left(  mq\right)  \left(  bc\right)  }^{ka}=\hat
{N}_{\left(  bc\right)  \left(  mq\right)  }^{ak}$ (see (\ref{eqnD83})).

We want to investigate the effect of eliminating the second class constraints
on the properties of the remaining canonical variables and on the PBs among
the functions constructed from them. $\hat{N}_{\left(  mq\right)  \left(
bc\right)  }^{ka}$ is given by (\ref{eqnD83}), and the explicit form of
$\hat{X}_{\left(  mq\right)  \left(  bc\right)  }^{ka}$ can be calculated, but
the inverse of $M$ can be defined for any $\hat{X}$. The explicit form of
$\hat{X}$ is needed only for calculation of DB among two translational
constraints, but we do not discuss this in this article. The inverse of such a
matrix can be immediately found if the inverse of the off-diagonal blocks is known%

\begin{equation}
\hat{N}_{\left(  mq\right)  \left(  bc\right)  }^{ka}\left(  \hat{N}%
^{-1}\right)  _{ax}^{\left(  bc\right)  \left(  yz\right)  }=\hat{I}_{x\left(
mq\right)  }^{k\left(  yz\right)  } \label{eqnD87}%
\end{equation}
where $\hat{I}_{x\left(  mq\right)  }^{k\left(  yz\right)  }$ is the
fundamental PB for antisymmetric and traceless canonical pair defined in
(\ref{eqnD68}). The inverse $\left(  \hat{N}^{-1}\right)  _{ax}^{\left(
bc\right)  \left(  yz\right)  }$ is%

\[
\left(  \hat{N}^{-1}\right)  _{ax}^{\left(  bc\right)  \left(  yz\right)
}=\frac{1}{4e}\left[  g_{ax}\left(  \gamma^{by}\gamma^{cz}-\gamma^{bz}%
\gamma^{cy}\right)  +\delta_{x}^{b}\left(  \gamma^{cz}\delta_{a}^{y}%
-\gamma^{cy}\delta_{a}^{z}\right)  -\delta_{x}^{c}\left(  \gamma^{bz}%
\delta_{a}^{y}-\gamma^{by}\delta_{a}^{z}\right)  \right.
\]

\begin{equation}
\left.  -\frac{2}{D-2}\left[  \delta_{a}^{b}\left(  \gamma^{cz}\delta_{x}%
^{y}-\gamma^{cy}\delta_{x}^{z}\right)  -\delta_{a}^{c}\left(  \gamma
^{bz}\delta_{x}^{y}-\gamma^{by}\delta_{x}^{z}\right)  \right]  \right]
.\label{eqnD88}%
\end{equation}
$\left(  \hat{N}^{-1}\right)  _{ax}^{\left(  bc\right)  \left(  yz\right)  }$
is also manifestly antisymmetric in $\left(  bc\right)  $ and $\left(
yz\right)  ,$ as well it is traceless in $ba$ and $ca,$ in $yx$ and $xz$ .The
following properties are useful%

\begin{equation}
\hat{I}\cdot\hat{I}=\hat{I},\text{ \ \ \ \ }\hat{I}\cdot\hat{N}=\hat{N},\text{
\ \ \ }\hat{I}\cdot\hat{N}^{-1}=\hat{N}^{-1}. \label{eqnD88a}%
\end{equation}

Equation (\ref{eqnD87}) and properties (\ref{eqnD88a}) allows us to find
$M^{-1}$%

\begin{equation}
M^{-1}=\left[
\begin{array}
[c]{cc}%
\hat{N}^{-1}\hat{X}\hat{N}^{-1} & -\hat{N}^{-1}\\
\hat{N}^{-1} & 0
\end{array}
\right]  ,\text{ \ \ \ }M\cdot M^{-1}=\left[
\begin{array}
[c]{cc}%
\hat{I} & 0\\
0 & \hat{I}%
\end{array}
\right]  . \label{eqnD89}%
\end{equation}

The canonical variables which remain after elimination of the pair ($\hat{\Pi
}_{\left(  mk\right)  }^{z},\hat{\Sigma}_{z}^{\ \left(  mk\right)  }$) have
fundamental PBs that are not affected, as can be easily checked. Substitution
of the solution of the secondary constraints (\ref{eqnD84}) into
(\ref{eqnD69c}) (Hamiltonian reduction) leads to%

\begin{equation}
H_{T}=\pi^{0\left(  \rho\right)  }\dot{e}_{0\left(  \rho\right)  }%
+\Pi^{0\left(  \alpha\beta\right)  }\dot{\omega}_{0\left(  \alpha\beta\right)
}-\omega_{0\left(  \alpha\beta\right)  }\chi^{0\left(  \alpha\beta\right)
}-e_{0\left(  \sigma\right)  }\chi^{0\left(  \sigma\right)  }\left(
\hat{\Sigma}_{z}^{\ \left(  mk\right)  }\text{ from Eq. }(\ref{eqnD84}%
)\right)  . \label{eqnD89a}%
\end{equation}
After substitution of $\hat{\Sigma}_{z}^{\ \left(  mk\right)  }$ this leads to
the same result as obtained in \cite{Report}. Darboux coordinates
significantly simplify the calculations.

The main goal of this paper is construction of Darboux coordinates for the EC
action in an independent of a particular dimension form. The direct approach
used in \cite{Report} made further calculations almost unmanageable, and the
simplification due to Darboux coordinates that shortens the calculations of
(\ref{eqnD89a}), gives us a hope of completing the Dirac procedure. We wish to
demonstrate its closure, the absence of tertiary constraints, and restore
gauge invariance. This result will be reported elsewhere. Here we just
demonstrate that with Darboux coordinates these calculations seems to become
manageable; and as an example, we consider the PB between rotational and
translational constraints. We argued in \cite{Report} that the known
invariance of the EC action under Lorentz rotation, Dirac's conjecture
\cite{Diracbook} and the Castellani algorithm \cite{Castellani} lead to the
necessity of having the PB among rotational and translational constraints
being exactly the same in all dimensions and given by the corresponding part
of the Poincar\'{e} algebra (\ref{eqnD8}).

Let us demonstrate that indeed in all dimensions ($D>2$) the PB among
translational and rotational constraints is the same and corresponds to the
Poincar\'{e} algebra (\ref{eqnD8}). This part of algebra, among the secondary
constraints $\left\{  \chi^{0\left(  \sigma\right)  },\chi^{0\left(  \mu
\nu\right)  }\right\}  ,$ as well as $\left\{  \chi^{0\left(  \sigma\right)
},\chi^{0\left(  \rho\right)  }\right\}  $, can be calculated using DBs, i.e.
avoiding substitution of $\hat{\Sigma}$ into the translational constraint
before calculating the PB (which is the longest part of such calculations) and
performing this substitution only after%

\[
\left\{  \chi^{0\left(  \sigma\right)  },\chi^{0\left(  \mu\nu\right)
}\right\}  _{DB}=
\]

\begin{equation}
\left\{  \chi^{0\left(  \sigma\right)  },\chi^{0\left(  \mu\nu\right)
}\right\}  -\left(
\begin{array}
[c]{cc}%
\left\{  \chi^{0\left(  \sigma\right)  },\hat{\Pi}_{~\left(  mq\right)  }%
^{k}\right\}   & \left\{  \chi^{0\left(  \sigma\right)  },\hat{\chi}_{~\left(
mq\right)  }^{k}\right\}
\end{array}
\right)  M^{-1}\left(
\begin{array}
[c]{c}%
\left\{  \hat{\Pi}_{~\left(  bc\right)  }^{a},\chi^{0\left(  \mu\nu\right)
}\right\}  \\
\left\{  \hat{\chi}_{~\left(  bc\right)  }^{a},\chi^{0\left(  \mu\nu\right)
}\right\}
\end{array}
\right)  ,\label{eqnD100}%
\end{equation}
where $\hat{\Sigma}$ and $\hat{\Pi}$ are the fundamental variables and the
only non-zero PBs are given in (\ref{eqnD67}-\ref{eqnD68}). (Note that only
the first step of reduction is performed in (\ref{eqn69b}).) After calculating
(\ref{eqnD100}) the solution of $\hat{\Sigma}$ is substituted that gives us
the final answer for $\left\{  \chi^{0\left(  \sigma\right)  },\chi^{0\left(
\mu\nu\right)  }\right\}  $. The advantage of this calculation is possibility
to demonstrate (and also single out) contributions in the first term of
(\ref{eqnD100}) that gives the corresponding part of the Poincar\'{e} algebra
almost manifestly and in compact form. First of all, we outline the idea of
such calculations. It is not difficult to demonstrate that the calculation of
the first PB in (\ref{eqnD100}) gives%

\begin{equation}
\left\{  \chi^{0\left(  \sigma\right)  },\chi^{0\left(  \mu\nu\right)
}\right\}  =\frac{1}{2}\tilde{\eta}^{\sigma\mu}\chi^{0\left(  \nu\right)
}\left(  \pi,e,\hat{\Sigma}\right)  -\frac{1}{2}\tilde{\eta}^{\sigma\nu}%
\chi^{0\left(  \mu\right)  }\left(  \pi,e,\hat{\Sigma}\right)  +R^{\sigma
\left(  \mu\nu\right)  }\left(  \hat{\Sigma},\pi,e\right)  . \label{eqnD101}%
\end{equation}

So, substitution of solution of $\hat{\Sigma}$ will not affect the first two
terms in (\ref{eqnD101}). What is left is to demonstrate that the remainder
$R^{\sigma\left(  \mu\nu\right)  }$, along with the second contribution in
(\ref{eqnD100}), after substitution of $\hat{\Sigma}$ gives zero which is a
long but straightforward calculation. It is easier to prove that
$R^{\sigma\left(  \mu\nu\right)  }$ is zero if we consider separately terms of
different nature, for example, all terms which are linear in momenta should
cancel independently of the rest of contributions. This allows us to break
these long and cumbersome calculations into smaller and independent pieces.

Let us outline the proof of (\ref{eqnD101}). We start calculations by
separating the translational constraint (\ref{eqnD74}) into contributions of
different order in $\hat{\Sigma}$%

\begin{equation}
\chi^{0\left(  \sigma\right)  }=\chi^{0\left(  \sigma\right)  }\left(
\hat{\Sigma}^{2}\right)  +\chi^{0\left(  \sigma\right)  }\left(  \hat{\Sigma
}^{1}\right)  +\chi^{0\left(  \sigma\right)  }\left(  \hat{\Sigma}^{0}\right)
. \label{eqnD102}%
\end{equation}

For the contribution quadratic $\hat{\Sigma}$ we almost immediately obtain the
exact expression (there are no contributions into the remainder in this order,
$R^{\sigma\left(  \mu\nu\right)  }\left(  \hat{\Sigma}^{2}\right)  =0$)%

\begin{equation}
\left\{  \chi^{0\left(  \sigma\right)  }\left(  \hat{\Sigma}^{2}\right)
,\chi^{0\left(  \mu\nu\right)  }\right\}  =\frac{1}{2}\tilde{\eta}^{\sigma\mu
}\chi^{0\left(  \nu\right)  }\left(  \hat{\Sigma}^{2}\right)  -\frac{1}%
{2}\tilde{\eta}^{\sigma\nu}\chi^{0\left(  \mu\right)  }\left(  \hat{\Sigma
}^{2}\right)  . \label{eqnD103}%
\end{equation}

Next contribution, linear in $\hat{\Sigma}$, is%

\[
\left\{  \chi^{0\left(  \sigma\right)  }\left(  \hat{\Sigma}\right)
,\chi^{0\left(  \mu\nu\right)  }\right\}  =
\]

\begin{equation}
\left(  e_{m\left(  \rho\right)  ,q}+e_{q}^{\left(  \gamma\right)  }%
\omega_{m\left(  \gamma\rho\right)  }\left(  \pi\right)  \right)  \hat{\Sigma
}_{k}^{\ \left(  mq\right)  }\left\{  2eA^{0\left(  \sigma\right)  k\left(
\rho\right)  },\frac{1}{2}\pi^{n\left(  \mu\right)  }e_{n}^{\left(
\nu\right)  }-\frac{1}{2}\pi^{n\left(  \nu\right)  }e_{n}^{\left(  \mu\right)
}\right\}  \label{eqnD104}%
\end{equation}

\[
+2eA^{0\left(  \sigma\right)  k\left(  \rho\right)  }\hat{\Sigma}%
_{k}^{\ \left(  mq\right)  }\left\{  \left(  e_{m\left(  \rho\right)
,q}+e_{q}^{\left(  \gamma\right)  }\omega_{m\left(  \gamma\rho\right)
}\left(  \pi\right)  \right)  ,\frac{1}{2}\pi^{n\left(  \mu\right)  }%
e_{n}^{\left(  \nu\right)  }-\frac{1}{2}\pi^{n\left(  \nu\right)  }%
e_{n}^{\left(  \mu\right)  }+eB^{n\left(  \rho\right)  m\left(  \mu\right)
0\left(  \nu\right)  }e_{n\left(  \rho\right)  ,m}\right\}  .
\]
Considering the PB in the first term of (\ref{eqnD104}) and using (see
(\ref{eqn41}) of Appendix A) we obtain%

\begin{equation}
\left\{  2eA^{0\left(  \sigma\right)  k\left(  \rho\right)  },\frac{1}{2}%
\pi^{n\left(  \mu\right)  }e_{n}^{\left(  \nu\right)  }-\frac{1}{2}%
\pi^{n\left(  \nu\right)  }e_{n}^{\left(  \mu\right)  }\right\}  =eB^{n\left(
\mu\right)  0\left(  \sigma\right)  k\left(  \rho\right)  }e_{n}^{\left(
\nu\right)  }-\left(  \mu\leftrightarrow\nu\right)  . \label{eqnD105}%
\end{equation}
Expanding $B$ in $n$ (see (\ref{eqn43}) of Appendix A) and contracting it with
$e_{n}^{\left(  \nu\right)  }$ we have%

\[
eB^{n\left(  \mu\right)  0\left(  \sigma\right)  k\left(  \rho\right)  }%
e_{n}^{\left(  \nu\right)  }=e\left(  \tilde{\eta}^{\mu\nu}-e^{0\left(
\mu\right)  }e_{0}^{\left(  \nu\right)  }\right)  A^{0\left(  \sigma\right)
k\left(  \rho\right)  }%
\]

\begin{equation}
+e\left(  \tilde{\eta}^{\sigma\nu}-e^{0\left(  \sigma\right)  }e_{0}^{\left(
\nu\right)  }\right)  A^{0\left(  \rho\right)  k\left(  \mu\right)  }+e\left(
\tilde{\eta}^{\rho\nu}-e^{0\left(  \rho\right)  }e_{0}^{\left(  \nu\right)
}\right)  A^{0\left(  \mu\right)  k\left(  \sigma\right)  }.\label{eqnD106}%
\end{equation}
Three terms in (\ref{eqnD106}) proportional to $e_{0}^{\left(  \nu\right)  }$
give us%

\[
-ee_{0}^{\left(  \nu\right)  }\left(  e^{0\left(  \mu\right)  }A^{0\left(
\sigma\right)  k\left(  \rho\right)  }+e^{0\left(  \sigma\right)  }A^{0\left(
\rho\right)  k\left(  \mu\right)  }+e^{0\left(  \rho\right)  }A^{0\left(
\mu\right)  k\left(  \sigma\right)  }\right)  .
\]

The expression in brackets exactly coincides with expansion of $B^{0\left(
\mu\right)  0\left(  \sigma\right)  k\left(  \rho\right)  }$ (see
(\ref{eqn43}) of Appendix A) which automatically equals zero because of
antisymmetry of $B$ in external indices (for details see Appendix A). After
antisymmetrization of the remaining terms of (\ref{eqnD105}) we have%

\[
\left\{  2eA^{0\left(  \sigma\right)  k\left(  \rho\right)  },\frac{1}{2}%
\pi^{n\left(  \mu\right)  }e_{n}^{\left(  \nu\right)  }-\frac{1}{2}%
\pi^{n\left(  \nu\right)  }e_{n}^{\left(  \mu\right)  }\right\}
\]

\[
=-e\tilde{\eta}^{\nu\sigma}A^{0\left(  \rho\right)  k\left(  \mu\right)
}-e\tilde{\eta}^{\nu\rho}A^{0\left(  \mu\right)  k\left(  \sigma\right)
}+e\tilde{\eta}^{\mu\sigma}A^{0\left(  \rho\right)  k\left(  \nu\right)
}+e\tilde{\eta}^{\mu\rho}A^{0\left(  \nu\right)  k\left(  \sigma\right)  }.
\]

The first and third terms contracted with the expression in front of the PB in
(\ref{eqnD104}) gives exactly two rotational constraints; and the rest of
terms, along with the second term in (\ref{eqnD104}), contribute to the
remainder. Finally,%

\begin{equation}
\left\{  \chi^{0\left(  \sigma\right)  }\left(  \hat{\Sigma}^{1}\right)
,\chi^{0\left(  \mu\nu\right)  }\right\}  =\frac{1}{2}\tilde{\eta}^{\sigma\mu
}\chi^{0\left(  \nu\right)  }\left(  \hat{\Sigma}^{1}\right)  -\frac{1}%
{2}\tilde{\eta}^{\sigma\nu}\chi^{0\left(  \mu\right)  }\left(  \hat{\Sigma
}^{1}\right)  +\ R^{\sigma\left(  \mu\nu\right)  }\left(  \hat{\Sigma}%
^{1}\right)  . \label{eqnD109}%
\end{equation}

Similarly one can demonstrate (using ABC properties) that in the last order
(zero order in $\hat{\Sigma}$ ) the same result as (\ref{eqnD109}) follows
(with the additional contributions into the remainder $R^{\sigma\left(  \mu
\nu\right)  }\left(  \hat{\Sigma}^{0}\right)  $) leading to (\ref{eqnD101}).
This part of calculation is simple and completes the proof of (\ref{eqnD101}).
To complete calculations, the remainder has to be considered together with the
second term of (\ref{eqnD100}), and the solution of $\hat{\Sigma}$ has to be
substituted at the end of calculations.

From the field content of $\chi^{0\left(  \mu\nu\right)  }$ (independence from
$\hat{\Pi}_{~\left(  bc\right)  }^{a}$) it follows that $\left\{  \hat{\Pi
}_{~\left(  bc\right)  }^{a},\chi^{0\left(  \mu\nu\right)  }\right\}  =0$.
Using this PB and the explicit form of $M^{-1}$ we obtain%

\[
\left(
\begin{array}
[c]{cc}%
\left\{  \chi^{0\left(  \sigma\right)  },\hat{\Pi}_{~\left(  mq\right)  }%
^{k}\right\}  & \left\{  \chi^{0\left(  \sigma\right)  },\hat{\chi}_{~\left(
mq\right)  }^{k}\right\}
\end{array}
\right)  M^{-1}\left(
\begin{array}
[c]{c}%
\left\{  \hat{\Pi}_{~\left(  bc\right)  }^{a},\chi^{0\left(  \mu\nu\right)
}\right\} \\
\left\{  \hat{\chi}_{~\left(  bc\right)  }^{a},\chi^{0\left(  \mu\nu\right)
}\right\}
\end{array}
\right)  =
\]

\begin{equation}
\left\{  \chi^{0\left(  \sigma\right)  },\hat{\Pi}_{~\left(  mq\right)  }%
^{k}\right\}  \left(  -\hat{N}^{-1}\right)  _{ka}^{\left(  mq\right)  \left(
bc\right)  }\left\{  \hat{\chi}_{~\left(  bc\right)  }^{a},\chi^{0\left(
\mu\nu\right)  }\right\}  . \label{eqnD110}%
\end{equation}

When calculating (\ref{eqnD110}), it is better to extract terms proportional
to $\hat{\chi}_{~\left(  bc\right)  }^{a}$, which after substitution of
$\hat{\Sigma}$ vanish and we are left with simple expressions%

\begin{equation}
\left\{  \chi^{0\left(  \sigma\right)  },\hat{\Pi}_{~\left(  mq\right)  }%
^{k}\right\}  =-2ee^{n\left(  \sigma\right)  }e^{0\left(  \beta\right)
}\left(  e_{p\left(  \beta\right)  ,d}+e_{d}^{\left(  \gamma\right)  }%
\omega_{p\left(  \gamma\beta\right)  }\left(  \pi\right)  \right)  \left\{
\hat{\Sigma}_{n}^{\ \left(  pd\right)  },\hat{\Pi}_{~\left(  mq\right)  }%
^{k}\right\}  \label{eqnD111}%
\end{equation}
and%

\begin{equation}
\left\{  \hat{\chi}_{~\left(  bc\right)  }^{a},\chi^{0\left(  \mu\nu\right)
}\right\}  =-e\left\{  \hat{D}_{\ \left(  bc\right)  }^{a},\chi^{0\left(
\mu\nu\right)  }\right\}  . \label{eqnD112}%
\end{equation}

Using (\ref{eqnD111}), (\ref{eqnD112}) together with (\ref{eqnD100}),
(\ref{eqnD88a}), antisymmetry and tracelessness of (\ref{eqnD88}) and
(\ref{eqnD80}), we obtain%

\begin{equation}
\left\{  \chi^{0\left(  \sigma\right)  },\chi^{0\left(  \mu\nu\right)
}\right\}  =\frac{1}{2}\tilde{\eta}^{\sigma\mu}\chi^{0\left(  \nu\right)
}-\frac{1}{2}\tilde{\eta}^{\sigma\nu}\chi^{0\left(  \mu\right)  }%
+\ R^{\sigma\left(  \mu\nu\right)  }\label{eqnD114}%
\end{equation}

\[
+4ee^{n\left(  \sigma\right)  }e^{0\left(  \beta\right)  }\left(  e_{p\left(
\beta\right)  ,b}+e_{b}^{\left(  \gamma\right)  }\omega_{p\left(  \gamma
\beta\right)  }\left(  \pi\right)  \right)  \left(  \hat{N}^{-1}\right)
_{na}^{\left(  pd\right)  \left(  bc\right)  }e\left\{  e^{a\left(
\alpha\right)  }e_{b\left(  \alpha\right)  ,c}+e^{a\left(  \alpha\right)
}e_{c}^{\left(  \rho\right)  }\omega_{b\left(  \rho\alpha\right)  }\left(
\pi\right)  ,\chi^{0\left(  \mu\nu\right)  }\right\}
\]

The most laborious part of calculation is a demonstration that the remainder
together with the last line of (\ref{eqnD114}) equals zero. We perform these
calculations by separating terms of different order in $\pi^{n\left(
\mu\right)  }$ that makes the analysis more manageable.

Using Darboux coordinates allows us to prove that the PB among rotational and
translational constraints also supports the Poincar\'{e} algebra in all
dimensions, $D>2$. Knowledge of this PB along with (\ref{eqnD6}) is sufficient
to restore rotational invariance in the Hamiltonian formulation of
Einstein-Cartan action by using the Castellani procedure. This result, as well
as calculation of the PB between two translational constraints and restoration
of translational invariance, will be reported elsewhere.

\section{Discussion}

Based on the results of the direct application of the Dirac procedure to the
first order Einstein-Cartan action \cite{Report}, we have constructed uniform
Darboux coordinates valid in all dimensions for which the first order
formulation exists; i.e. when it is equivalent to the second order EC
action$\ $($D>2$). In particular these uniform Darboux coordinates guarantee
equivalence and allow one to check the $D=3$ limit \cite{3D} at all stages of
calculations in dimensions $D>3$. Considerable simplification occurs when we
use Darboux coordinates and it is explicitly demonstrated by obtaining the
Hamiltonian formulation in a few lines compared with the direct and cumbersome
calculations \cite{Report} considered previously. However, we have to
emphasize that the preliminary Hamiltonian analysis is indispensable for the
construction of Darboux coordinates, which preserve equivalence with the
original action. An arbitrary change of variables at the Lagrangian level for
singular Lagrangians is an ambiguous operation because it might correspond to
a non-canonical change of variables at the Hamiltonian level. For singular
Lagrangians the invertability of transformations (redefinition of fields) from
one set of variables to another is not a sufficient condition to preserve
equivalence \cite{GGT}; and one particular example is considered in the end of
Section III (see (\ref{eqnD56})). These \textquotedblleft Darboux
coordinates\textquotedblright, (\ref{eqnD56}), despite separating variables in
the same way, do not preserve the $D=3$ limit, lead to results which are
different from those found by the direct analysis and destroy equivalence. Our
Darboux transformations, (\ref{eqnD13}), do not suffer such an ambiguity
because they are based on the preliminary Hamiltonian analysis. In other
words, our transformation separates variables in the same way that the
Hamiltonian reduction does. This makes this transformation unique and
preserves equivalence with the original action as well as the equivalence of
results for the Lagrangian and Hamiltonian formulations.

To answer the question about possible modifications of the Poincar\'{e}
algebra of PBs among the secondary constraints of the EC Hamiltonian in
dimensions $D>3$, we need to complete the calculations of PBs. In particular,
if $\left\{  \chi^{0\left(  \sigma\right)  },\chi^{0\left(  \mu\nu\right)
}\right\}  =\frac{1}{2}\tilde{\eta}^{\sigma\mu}\chi^{0\left(  \nu\right)
}-\frac{1}{2}\tilde{\eta}^{\sigma\nu}\chi^{0\left(  \mu\right)  }$ (and there
is a strong indication that this is the case) and $\left\{  \chi^{0\left(
\alpha\right)  },\chi^{0\left(  \beta\right)  }\right\}  =0,$ then the N-bein
gravity is the Poincar\'{e} gauge theory in all dimensions and the $D=3$ case
is not special at all. Note (we discussed this in \cite{Report}) that in
higher dimensions, the constraints are much more complicated and having the
same algebra does not mean that the gauge transformations must be exactly the
same as for $D=3$. If $\left\{  \chi^{0\left(  \alpha\right)  },\chi^{0\left(
\beta\right)  }\right\}  \neq0$, but proportional to secondary first class
constraints, we still have closure of the Dirac procedure, all constraints are
first class, the gauge generators can be found and the gauge transformations
can be restored. In this case, N-bein gravity for $D>3$ is the gauge theory,
but with the modified Poincar\'{e} algebra. For example, if (the most general case)%

\begin{equation}
\left\{  \chi^{0\left(  \alpha\right)  },\chi^{0\left(  \beta\right)
}\right\}  =\tilde{F}_{\left(  \mu\nu\right)  }^{\left(  \alpha\beta\right)
}\chi^{0\left(  \mu\nu\right)  }+\tilde{M}_{\nu}^{\left(  \alpha\beta\right)
}\chi^{0\left(  \nu\right)  }\label{D1}%
\end{equation}
with structure functions $\tilde{F}_{\left(  \mu\nu\right)  }^{\left(
\alpha\beta\right)  }$ and $\tilde{M}_{\nu}^{\left(  \alpha\beta\right)  }$,
which are $\tilde{F}_{\left(  \mu\nu\right)  }^{\left(  \alpha\beta\right)
}=0$ and $\tilde{M}_{\nu}^{\left(  \alpha\beta\right)  }=0$ when $D=3$, then
one can say that in all dimensions the EC theory is a gauge theory with a
generalized Poincar\'{e} algebra among secondary first class constraints that
degenerates into the true Poincar\'{e} algebra when $D=3$. A similar result as
(\ref{D1}) has been known for a long time and was presented in \cite{Trautman,
Hehl-1, Leclerc}; but it was not obtained using the Hamiltonian procedure and
it was written for generators, not for the PBs of constraints. The complete
Hamiltonian analysis will show whether the algebra among constraints is
Poincar\'{e} or modified Poincar\'{e}.

The exact form of algebra of secondary constraints is important; but we
already have enough evidence to make a conclusion about the gauge invariance
of the EC action. It is certain that \textit{gauge} invariance is the
translation and Lorentz rotation in the internal space and that diffeomorphism
(either spatial or full) is not a \textit{gauge invariance} of N-bein gravity
generated by first class constraints. This conclusion is based on the
following arguments.

The parameters characterizing the gauge transformations are defined by the
tensorial nature of primary first class constraints. Both of them,
$\Pi^{0\left(  \alpha\beta\right)  }$ and $\pi^{0\left(  \sigma\right)  }$,
have internal indices, so do the corresponding gauge parameters $r_{\left(
\alpha\beta\right)  }$ and $t_{\left(  \sigma\right)  }$ (see Section 5 of
\cite{Myths-2} for more details). This gauge symmetry corresponds to rotation
and translation in the tangent space. The gauge parameter of diffeomorphism,
$\xi^{\mu}$, has an external index, which can be accommodated only if the
corresponding primary first class constraint has also an external free index.
This does not happen in the case of N-bein gravity if the Dirac procedure is
performed correctly and a non-canonical change of variable is not made (see
\cite{Myths-2}). Formulations that claim to have the \textquotedblleft spatial
diffeomorphism constraint\textquotedblright\ or any further consideration
based on such a constraint\ \cite{Thiemann, Gambini} for tetrad gravity is the
product of non-canonical change of variables, which has the same origin as in
metric gravity.\footnote{The loss of full diffeomorphism invariance due to a
non-canonical change of variables in metric gravity was discussed in
\cite{Myths, KKRV, FKK}. A lapse with canonicity leads to big (or rather a
devastating) shift from covariant General Relativity (Einstein-Hilbert and
Einstein-Cartan actions) to some non-covariant models like \textquotedblleft
geometrodynamics\textquotedblright\ for metric \ gravity and inspired by it
non-covariant models for tetrads (see Section V of \cite{Myths-2} for
discussion on this topic). This, of course, propagates into further analysis
(i.e. quantization) of these models.} Gauge invariance is a unique
characteristic of a singular system and follows from its unique constraint
structure, i.e. it is derivable from the first class constraints in accordance
with Dirac's conjecture \cite{Diracbook} and using the Castellani algorithm
\cite{Castellani} (the only true algorithm to restore gauge invariance
\cite{Myths-2}). Gauge invariance is unique, but it does not presume the
absence of additional symmetries in the action. In \cite{HT} such symmetries,
which are not derivable from constraints, are called \textquotedblleft
trivial\textquotedblright\footnote{In Section 3.1.5. of \cite{HT}
\textquotedblleft trivial gauge transformations\textquotedblright\ are defined
and all transformations are classified by using the Hamiltonian method and
\textquotedblleft the transformations are of no physical significance because
in the Hamiltonian formalism they are not generated by a
constraint\textquotedblright.\ } but this name seems to us to be a little bit
confusing as e.g. the non-gauge symmetry of the EC action, diffeomorphism, can
hardly be called \textquotedblleft trivial\textquotedblright. It is more
preferable, without introducing any new terminology, to just have
\textquotedblleft symmetries of the action\textquotedblright\ (there could be
many)\ and \textquotedblleft a gauge symmetry\textquotedblright(a unique one)
that follows from the Hamiltonian analysis or from basic differential
identities at the Lagrangian level \cite{Trans}.

The Hamiltonian analysis allows us to single out \textquotedblleft what is a
gauge symmetry and what is not\textquotedblright\ \cite{Matschull}. In
\cite{Trans} we showed, using differential identities, that the translation in
the internal space is an invariance of the EC action. This fact has long been
known; such transformations were written in \cite{Hehl-1, Leclerc}, and are
exactly the same as we obtained. This makes the common statement that
\textquotedblleft translation is not invariance\textquotedblright\ absolutely
groundless and somewhat mysterious. In \cite{Trans} we also argue that two
invariances, translation in the internal space and diffeomorphism, cannot
coexist (simultaneously present) as \textit{gauge} invariances in Hamiltonian
formulations as the number of first class constraints needed to generate both
of them would lead to a negative number of degrees of freedom (for relation
between the number of constraints and degrees of freedom see \cite{HTZ}). The
only possible way to reconcile these two symmetries, as we stated in
\cite{Trans}, is that there exists a canonical transformation that converts
our constraints (\ref{eqnD89a}) into a different set of constraints which
support diffeomorphism. In this article we argue that this is impossible and
such a canonical transformation does not exist. Let us look at this from the
Hamiltonian and Lagrangian points of view.

From the Hamiltonian point of view, the known canonical transformations for
the first and second order Einstein-Hilbert actions \cite{GKK, FKK, Myths-2}
always preserve the form-invariance of the constraint algebra that would be
destroyed by any transformation that changes the tensorial character of the
primary constraints needed to have diffeomorphism as a \textit{gauge}
invariance of the EC action (as the gauge parameter of diffeomorphism
$\xi^{\mu}$ is a true vector; for more details see \cite{Myths-2}). In
addition, in formulations which are related by a canonical transformation,
constraints are different, but the gauge transformations are the same and for
new and old fields can be obtained one from another without any need for a
field dependent redefinition of gauge parameters (such a field dependent
redefinition is an indication of having a non-canonical change of variables).
The gauge parameters which are responsible for diffeomorphism, $\xi^{\mu}$,
and translation, $t_{\left(  \sigma\right)  }$, cannot be related without
involving fields as they have a different tensorial dimension. Therefore, the
field dependent redefinition of parameters is needed. Consequently, there is
no canonical transformation between such formulations which give
diffeomorphism and translation as gauge symmetries. And so, the Hamiltonian
formulation with the constraints that would produce diffeomorphism invariance
is not equivalent to the Hamiltonian formulation with translation in the
tangent space as a gauge symmetry.

From the Lagrangian point of view, using the 16 components of the tetrads (in
the $D=4$ case) we can restore the 10 components of the metric tensor but
\textquotedblleft not vice versa\textquotedblright\ \cite{Einstein}. Tetrads
are \textquotedblleft world\textquotedblright\ vectors and are invariant under
diffeomorphism, as any vector or tensor does in a generally covariant theory.
From the diffeomorphism invariance of tetrads we can derive invariance under
diffeomorphism for any combination of tetrads, in particular, for $e_{\mu
}^{\left(  \alpha\right)  }e_{\nu\left(  \alpha\right)  }=$ $g_{\mu\nu}$ we obtain%

\[
\delta_{diff}\left(  e_{\mu}^{\left(  \alpha\right)  }e_{\nu\left(
\alpha\right)  }\right)  =e_{\mu}^{\left(  \alpha\right)  }\left(
-e_{\rho\left(  \alpha\right)  }\xi_{,\nu}^{\rho}-e_{\nu\left(  \alpha\right)
,\rho}\xi^{\rho}\right)  +\left(  -e_{\rho}^{\left(  \alpha\right)  }\xi
_{,\mu}^{\rho}-e_{\mu,\rho}^{\left(  \alpha\right)  }\xi^{\rho}\right)
e_{\nu\left(  \alpha\right)  }%
\]

\begin{equation}
=-\xi_{\mu,\nu}+g_{\rho\mu,\nu}\xi^{\rho}-\xi_{\nu,\mu}+g_{\rho\nu,\mu}%
\xi^{\rho}-g_{\mu\nu,\rho}\xi^{\rho}=\delta_{diff}g_{\mu\nu}. \label{eqn-diff}%
\end{equation}
Note that in this case there is no need for a field dependent redefinition of
gauge parameters.

We can also perform the inverse operation: from the diffeomorphism of the
metric tensor (which is the gauge invariance of Einstein-Hilbert (EH) action
\cite{KKRV, Myths, FKK, Myths-2}) we can derive the diffeomorphism of tetrads
(more details of this derivation and discussion about gauge symmetries of the
metric tensor and tetrads are given in \cite{3D}). But we cannot obtain
Lorentz or translational invariances in the tangent space of the EC action
from the diffeomorphism of the metric tensor. The reason for this is simply
that the EC and EH actions are not equivalent and neither are the
corresponding Hamiltonians: they have a different number of phase-space
variables, different constraints, PB algebras, tensorial dimension of primary
constraints and different gauge invariances.

In the Lagrangian formalism, if we perform a change of variables that keep the
equivalence of two formulations, build differential identities and restore
invariance of each formulation, then the invariance of one formulation must be
derivable from the invariance of another, using the same original redefinition
of fields and without redefinition of parameters. If a field dependent
redefinition of parameters is needed, then the change of variables that was
performed is not canonical at the Hamiltonian level. Hence, such a change of
variables destroy equivalence also at the Lagrangian level. As an illustration
of this, we again compare metric General Relativity (GR) and ADM gravity. For
metric GR the Hamiltonian \cite{KKRV, Myths} and Lagrangian \cite{Samanta}
methods give the same gauge transformation, diffeomorphism. For the ADM
gravity the Hamiltonian and Lagrangian methods also produce the same
invariance, which is different from diffeomorphism (compare  \cite{Saha} and
\cite{Banerjee}). This is consistent with the fact that the Hamiltonian and
Lagrangian approaches give equivalent descriptions of the same system.
However, it is clear from \cite{Saha} and \cite{Banerjee}, that the gauge
transformations of ADM gravity do not coincide with diffeomorphism. It is not
a surprise as ADM gravity is not equivalent to GR (see \cite{Myths, Myths-2}).
Only after a field dependent redefinition of parameters is performed, is it
possible to find \textquotedblleft equivalence between diffeomorphism and
gauge transformations\textquotedblright\ of ADM gravity \cite{Saha}. The same
redefinition is also needed at the Lagrangian level \cite{Banerjee} which,
according to the authors, demonstrates \textquotedblleft the equivalence
between the gauge and diff parameters by devising of the one to one
mapping\textquotedblright. The same is true for the EC gravity: the field
dependent redefinition of parameters is needed to relate translation and
diffeomorphism, so only non-canonical transformations can relate two such
Hamiltonian formulations.

An additional argument to support our point of view is related to differential
identities from which the invariances of a Lagrangian can be found. As we
showed in \cite{Trans}, considering the EC Lagrangian as an example, all
differential identities can be constructed from a few basic differential
identities. One out of many identities leading to invariances of the EC action
a \textit{gauge} identity can be singled out using the following arguments.
Differential identities leading to \textit{gauge} invariances for known
theories are always the simplest: they are built \textit{starting} by
contracting derivatives ($\partial_{\mu}$) with the Euler derivatives ($E$).
For example: for Maxwell theory it is $\partial_{\mu}E^{\mu}$, for Yang-Mills
- $\partial_{\mu}E^{\mu\left(  a\right)  }$, for the second order metric GR-
$\partial_{\mu}E^{\mu\nu}$. Variation of the action with respect to the
fundamental (basic) fields of a theory defines a tensorial character of a
differential identity. For the first order EC action the Euler derivatives are
$E^{\mu\left(  \alpha\right)  }=\frac{\delta L_{EC}}{\delta e_{\mu\left(
\alpha\right)  }}$ and $E^{\mu\left(  \alpha\beta\right)  }=\frac{\delta
L_{EC}}{\delta\omega_{\mu\left(  \alpha\beta\right)  }}$ \cite{Trans}. Thus,
the basic (the most fundamental) differential identities can be constructed
starting from $\partial_{\mu}E^{\mu\left(  \alpha\right)  }$ and
$\partial_{\mu}E^{\mu\left(  \alpha\beta\right)  }$ that lead to the following
identities $I^{\left(  \alpha\right)  }=\partial_{\mu}E^{\mu\left(
\alpha\right)  }+...$ and $I^{\left(  \alpha\beta\right)  }=\partial_{\mu
}E^{\mu\left(  \alpha\beta\right)  }+...$(see \cite{Trans}) which give rise to
the \textit{translational and rotational invariances in the tangent space}
\cite{Trautman}. The Hamiltonian method applied to singular systems (the Dirac
procedure) always leads to first class constraints that allow the restoration
of the \textit{gauge} invariance. The Hamiltonian formulation of the EC action
leads to the first class constraints with the PB algebra that describe
internal translation and rotation. This is clear from the first steps of the
Dirac procedure and the tensorial character of the primary first class
constraints \cite{3D, Report}. The same result, translational and rotational
invariances, also follows from the analysis of basic differential identities
at the Lagrangian level \cite{Trans}.

From the equivalence of the descriptions given by the Hamiltonian and
Lagrangian formulations, we can conclude that diffeomorphism is \textit{not a
gauge symmetry} derivable from the first class constraints of the EC action
and its Hamiltonian. \textit{Gauge} symmetries are \textit{translation and
rotation in the tangent space,} which makes the EC theory very similar to the
Yang-Mills theory. This is a classical result which, in particular, is related
to the classical background of Loop Quantum Gravity (LQG). One can say that
this is a classical question and it has nothing to do with quantum issues of
LQG, e.g. according to \cite{GRC} \textquotedblleft at the current stage, with
the author's [ours] focus on classical questions while the debated issues of
loop quantum gravity arise after quantization, their [our] criticism is rather
empty\textquotedblright. The logic of this statement is simple \footnote{This
logic cannot be debated in a scientific journal (debates cannot be published)
perhaps because of authorships of the statement \cite{GRC}.\ }: it does not
matter what a model is quantized, it does not matter whether the result is
correct or not, because one can always debate the result obtained after
quantization. In our opinion, if on the classical level the true gauge
invariance was missing because the Hamiltonian formulation produces something
different (e.g. \textquotedblleft spatial diffeomorphism\textquotedblright\ of
LQG) then the formulation, which is used for quantization, is
\textit{classically} not equivalent to the original Einstein-Cartan action.
With such a discrepancy at a very basic level, to debate the results that
arise after quantization of a classically different formulation is rather an
empty exercise (or a waste of time) for anyone who is interested in quantizing
Einstein gravity. Any quantization starts from a classical action. What
classical action is quantized in LQG? Contrary to the undebatable view
expressed in \cite{GRC}, there is a different opinion that coincides with our
point of view. In the conclusion to the review article on LQG of Nicolai et
al. \cite{Nicolai} one can read \textquotedblleft...despite the optimism
prevalent in many other reviews, more attention should be paid to basic
aspects and unresolved problems of the theory\textquotedblleft\ and a few
lines later the authors repeat again \textquotedblright...there are still too
many problems at a basic level that need to be addressed and
resolved\textquotedblright. We hope that our results provide enough arguments
to re-focus attention to the basic, classical, issues from \textquotedblleft
the debated issues of loop quantum gravity\textquotedblright\ that
\textquotedblleft arise after quantization\textquotedblright.

The answer to Matshull's question \cite{Matschull} \textquotedblleft what is a
gauge symmetry and what is not\textquotedblright\ is given for the EC theory
in any dimension. This is enough for covariant methods of quantization (based
on the Lagrangian formalism). However, for canonical quantization, the
Hamiltonian formulation of the EC action has to be completed. The unique
Darboux coordinates, the main subject of this article, provide a great
simplification of the Hamiltonian analysis. The modification of the algebra of
PBs among first class secondary constraints (see (\ref{D1})) compared to the
$D=3$ case and the restoration of gauge invariance (as consistency check) from
the constraint structure of the EC Hamiltonian using the Castellani algorithm
\cite{Castellani} is under our current investigation and will be reported elsewhere.

\bigskip\textbf{ACKNOWLEDGMENTS}

The authors are grateful to A.M. Frolov, P.G. Komorowski, D.G.C. McKeon, and
A.V. Zvelindovsky for numerous discussions and suggestions. The partial
support of The Huron University College Faculty of Arts and Social Science
Research Grant Fund is greatly acknowledged.

\appendix

\section{ $ABC$ properties}

Here we collect properties of the $ABC$ functions that were introduced in
considering the Hamiltonian formulation of N-bein gravity \cite{Report, 3D}.
They also turn out to be very useful in the Lagrangian formalism \cite{Trans}.

These functions are generated by consecutive variation of the N-bein density%

\begin{equation}
\frac{\delta}{\delta e_{\nu\left(  \beta\right)  }}\left(  ee^{\mu\left(
\alpha\right)  }\right)  =e\left(  e^{\mu\left(  \alpha\right)  }e^{\nu\left(
\beta\right)  }-e^{\mu\left(  \beta\right)  }e^{\nu\left(  \alpha\right)
}\right)  =eA^{\mu\left(  \alpha\right)  \nu\left(  \beta\right)  },
\label{eqn40}%
\end{equation}

\begin{equation}
\frac{\delta}{\delta e_{\lambda\left(  \gamma\right)  }}\left(  eA^{\mu\left(
\alpha\right)  \nu\left(  \beta\right)  }\right)  =eB^{\lambda\left(
\gamma\right)  \mu\left(  \alpha\right)  \nu\left(  \beta\right)  },\text{ \ }
\label{eqn41}%
\end{equation}

\begin{equation}
\frac{\delta}{\delta e_{\tau\left(  \sigma\right)  }}\left(  eB^{\lambda
\left(  \gamma\right)  \mu\left(  \alpha\right)  \nu\left(  \beta\right)
}\right)  =eC^{\tau\left(  \sigma\right)  \lambda\left(  \gamma\right)
\mu\left(  \alpha\right)  \nu\left(  \beta\right)  },\quad\label{eqn41a}%
\end{equation}

\begin{equation}
\frac{\delta}{\delta e_{\varepsilon\left(  \rho\right)  }}\left(
eC^{\tau\left(  \sigma\right)  \lambda\left(  \gamma\right)  \mu\left(
\alpha\right)  \nu\left(  \beta\right)  }\right)  =eD^{\varepsilon\left(
\rho\right)  \tau\left(  \sigma\right)  \lambda\left(  \gamma\right)
\mu\left(  \alpha\right)  \nu\left(  \beta\right)  },\quad... \label{eqn41b}%
\end{equation}

The first important property of these density functions is their total
antisymmetry: interchange of two indices of the same nature (internal or
external), e.g.%

\begin{equation}
A^{\nu\left(  \beta\right)  \mu\left(  \alpha\right)  }=-A^{\nu\left(
\alpha\right)  \mu\left(  \beta\right)  }=-A^{\mu\left(  \beta\right)
\nu\left(  \alpha\right)  }\label{eqn42}%
\end{equation}
with the same being valid for $B$, $C$, $D,$ etc. In particular, the presence
of two equal indices of the same nature (both internal or both external) makes
the functions $A$, $B$, etc. equal zero.

The second important property is their expansion using an external index%

\begin{equation}
B^{\tau\left(  \rho\right)  \mu\left(  \alpha\right)  \nu\left(  \beta\right)
}=e^{\tau\left(  \rho\right)  }A^{\mu\left(  \alpha\right)  \nu\left(
\beta\right)  }+e^{\tau\left(  \alpha\right)  }A^{\mu\left(  \beta\right)
\nu\left(  \rho\right)  }+e^{\tau\left(  \beta\right)  }A^{\mu\left(
\rho\right)  \nu\left(  \alpha\right)  }, \label{eqn43}%
\end{equation}

\begin{equation}
C^{\tau\left(  \rho\right)  \lambda\left(  \sigma\right)  \mu\left(
\alpha\right)  \nu\left(  \beta\right)  }=e^{\tau\left(  \rho\right)
}B^{\lambda\left(  \sigma\right)  \mu\left(  \alpha\right)  \nu\left(
\beta\right)  }-e^{\tau\left(  \sigma\right)  }B^{\lambda\left(
\alpha\right)  \mu\left(  \beta\right)  \nu\left(  \rho\right)  }%
+e^{\tau\left(  \alpha\right)  }B^{\lambda\left(  \beta\right)  \mu\left(
\rho\right)  \nu\left(  \sigma\right)  }-e^{\tau\left(  \beta\right)
}B^{\lambda\left(  \rho\right)  \mu\left(  \sigma\right)  \nu\left(
\alpha\right)  } \label{eqn44}%
\end{equation}
or an internal index%

\begin{equation}
B^{\tau\left(  \rho\right)  \mu\left(  \alpha\right)  \nu\left(  \beta\right)
}=e^{\tau\left(  \rho\right)  }A^{\mu\left(  \alpha\right)  \nu\left(
\beta\right)  }+e^{\mu\left(  \rho\right)  }A^{\nu\left(  \alpha\right)
\tau\left(  \beta\right)  }+e^{\nu\left(  \rho\right)  }A^{\tau\left(
\alpha\right)  \mu\left(  \beta\right)  }, \label{eqn45}%
\end{equation}

\begin{equation}
C^{\tau\left(  \rho\right)  \lambda\left(  \sigma\right)  \mu\left(
\alpha\right)  \nu\left(  \beta\right)  }=e^{\tau\left(  \rho\right)
}B^{\lambda\left(  \sigma\right)  \mu\left(  \alpha\right)  \nu\left(
\beta\right)  }-e^{\lambda\left(  \rho\right)  }B^{\mu\left(  \sigma\right)
\nu\left(  \alpha\right)  \tau\left(  \beta\right)  }+e^{\mu\left(
\rho\right)  }B^{\nu\left(  \sigma\right)  \tau\left(  \alpha\right)
\lambda\left(  \beta\right)  }-e^{\nu\left(  \rho\right)  }B^{\tau\left(
\sigma\right)  \lambda\left(  \alpha\right)  \mu\left(  \beta\right)  }.
\label{eqn46}%
\end{equation}

The third property involves their derivatives%

\begin{equation}
\left(  eA^{\nu\left(  \beta\right)  \mu\left(  \alpha\right)  }\right)
,_{\sigma}=\frac{\delta}{\delta e_{\lambda\left(  \gamma\right)  }}\left(
eA^{\nu\left(  \beta\right)  \mu\left(  \alpha\right)  }\right)
e_{\lambda\left(  \gamma\right)  ,\sigma}=eB^{\lambda\left(  \gamma\right)
\nu\left(  \beta\right)  \mu\left(  \alpha\right)  }e_{\lambda\left(
\gamma\right)  ,\sigma}~,\label{eqn47}%
\end{equation}

\begin{equation}
\left(  eB^{\tau\left(  \rho\right)  \nu\left(  \beta\right)  \mu\left(
\alpha\right)  }\right)  ,_{\sigma}=\frac{\delta}{\delta e_{\lambda\left(
\gamma\right)  }}\left(  eB^{\tau\left(  \rho\right)  \nu\left(  \beta\right)
\mu\left(  \alpha\right)  }\right)  e_{\lambda\left(  \gamma\right)  ,\sigma
}=eC^{\tau\left(  \rho\right)  \lambda\left(  \gamma\right)  \nu\left(
\beta\right)  \mu\left(  \alpha\right)  }e_{\tau\left(  \rho\right)  ,\sigma
}~.\label{eqn46a}%
\end{equation}

We also use the contraction of $B^{\tau\left(  \rho\right)  \mu\left(
\alpha\right)  \nu\left(  \beta\right)  }$ (\ref{eqn43}) with a covariant
$e_{\tau\left(  \lambda\right)  }:$%

\begin{equation}
e_{\tau\left(  \lambda\right)  }B^{\tau\left(  \rho\right)  \mu\left(
\alpha\right)  \nu\left(  \beta\right)  }=\tilde{\delta}_{\lambda}^{\rho
}A^{\mu\left(  \alpha\right)  \nu\left(  \beta\right)  }+\tilde{\delta
}_{\lambda}^{\alpha}A^{\mu\left(  \beta\right)  \nu\left(  \rho\right)
}+\tilde{\delta}_{\lambda}^{\beta}A^{\mu\left(  \rho\right)  \nu\left(
\alpha\right)  }. \label{eqn42a}%
\end{equation}
The above properties considerably simplify the calculations.

\section{Solution of the equation of motion for $\hat{\Sigma}$}

To eliminate the $\hat{\Sigma}$ field in the course of the Lagrangian or
Hamiltonian reduction, we perform variation of (\ref{eqnD61}) or
(\ref{eqnD66}) with respect to $\hat{\Sigma}$ and solve this equation for
$\hat{\Sigma}$. The corresponding part of the Lagrangian (Hamiltonian),
quadratic and linear in $\hat{\Sigma}$, after changing dummy indices and
performing some contractions is%

\begin{equation}
L\left(  \hat{\Sigma}\right)  =eg_{qp}\hat{\Sigma}_{k}^{\ \left(  mq\right)
}\hat{\Sigma}_{m}^{\ \left(  kp\right)  }+2e\hat{\Sigma}_{m}^{\ \left(
pq\right)  }\hat{D}_{\ pq}^{m} \label{A3-00}%
\end{equation}
where $g_{qp}=e_{q\left(  \alpha\right)  }e_{p}^{\left(  \alpha\right)  }$ and%

\begin{equation}
\hat{D}_{\ pq}^{m}=e^{m\left(  \beta\right)  }e_{p\left(  \beta\right)
,q}+e^{m\left(  \beta\right)  }e_{q}^{\left(  \gamma\right)  }N_{p\left(
\gamma\beta\right)  0n\left(  \sigma\right)  }\pi^{n\left(  \sigma\right)  }.
\label{A3-1}%
\end{equation}

Note that there are no symmetries in this expression for $pq$ indices (e.g.,
$\hat{D}_{\ pq}^{m}\neq-\hat{D}_{\ qp}^{m}$, $\hat{D}_{\ pq}^{m}\neq\hat
{D}_{\ qp}^{m}$), which is clear from its explicit form. Of course, we can do
further contraction in the second term of (\ref{A3-1}); but to find the
solution it is not necessary as it can be expressed in terms of the whole
$\hat{D}_{\ pq}^{m}$ and the separation of it into contributions with momenta
and spatial derivatives of covariant N-bein is sufficient on this stage and
keep the expressions in compact form.

Variations of a traceless antisymmetric field (see the fundamental PB
(\ref{eqnD68})) is%

\begin{equation}
\frac{\delta\hat{\Sigma}_{k}^{\ \left(  mq\right)  }}{\delta\hat{\Sigma}%
_{x}^{\ \left(  yz\right)  }}=\delta_{k}^{x}\hat{\Delta}_{\left(  yz\right)
}^{\left(  mq\right)  }-\frac{1}{D-2}\left[  \delta_{y}^{x}\hat{\Delta
}_{\left(  kz\right)  }^{\left(  mq\right)  }-\delta_{z}^{x}\hat{\Delta
}_{\left(  ky\right)  }^{\left(  mq\right)  }\right]  .\label{A3-2}%
\end{equation}
It is clear from (\ref{A3-2}) that this expression is antisymmetric in $mq$
and $yz$ and equals zero if the traces of $\hat{\Sigma}_{k}^{\ \left(
mq\right)  }$ or $\hat{\Sigma}_{x}^{\ \left(  yz\right)  }$ are taken.
Variation of (\ref{A3-00}) gives%

\begin{equation}
\hat{\Sigma}_{y}^{\ \left(  px\right)  }g_{zp}-\hat{\Sigma}_{z}^{\ \left(
px\right)  }g_{yp}=\hat{D}_{\ \left(  yz\right)  }^{x} \label{A3-3}%
\end{equation}
where%

\begin{equation}
\hat{D}_{\ \left(  yz\right)  }^{x}=\hat{D}_{\ yz}^{x}-\hat{D}_{\ zy}%
^{x}-\frac{1}{D-2}\left[  \delta_{y}^{x}\left(  \hat{D}_{\ mz}^{m}-\hat
{D}_{\ zm}^{m}\right)  -\delta_{z}^{x}\left(  \hat{D}_{\ my}^{m}-\hat
{D}_{\ ym}^{m}\right)  \right]  , \label{A3-4}%
\end{equation}
which is manifestly antisymmetric and traceless as it should be after
variation with respect to the field with such properties.

To solve equation (\ref{A3-3}) we use Einstein's permutation \cite{Einstein-2}%
. To do this we must have three indices of the same nature, either all
external or all internal, and in the same position, covariant or
contravariant. We can achieve this by contracting (\ref{A3-3}) with $g_{wx}$%

\begin{equation}
g_{wx}\hat{\Sigma}_{y}^{\ \left(  px\right)  }g_{zp}-g_{wx}\hat{\Sigma}%
_{z}^{\ \left(  px\right)  }g_{yp}=g_{wx}\hat{D}_{\ \left(  yz\right)  }^{x}.
\label{A3-5}%
\end{equation}

Now we have combinations with three free external indices in covariant
position and can use the permutation $\left(  wyz\right)  +\left(  yzw\right)
-\left(  zwy\right)  $ that gives us%

\begin{equation}
2g_{yx}\hat{\Sigma}_{z}^{\ \left(  px\right)  }g_{wp}=g_{wx}\hat{D}_{\ \left(
yz\right)  }^{x}+g_{yx}\hat{D}_{\ \left(  zw\right)  }^{x}-g_{zx}\hat
{D}_{\ \left(  wy\right)  }^{x}. \label{A3-6}%
\end{equation}

To find explicitly $\hat{\Sigma}_{z}^{\ \left(  px\right)  }$, we have to use
the Dirac inverse $\gamma^{km}=g^{km}-\frac{g^{0m}g^{0k}}{g^{00}}$. (We
repeat, it is not a new variable, but a short-hand notation for a particular
combination of N-bein fields). After contracting (\ref{A3-6}) with
$\gamma^{ky}\gamma^{mw}$ we obtain the solution%

\begin{equation}
\hat{\Sigma}_{z}^{\ \left(  mk\right)  }=\frac{1}{2}\left(  \gamma^{kb}\hat
{D}_{\ \left(  bz\right)  }^{m}-\gamma^{mb}\hat{D}_{\ \left(  bz\right)  }%
^{k}-\gamma^{ky}\gamma^{mw}g_{zx}\hat{D}_{\ \left(  wy\right)  }^{x}\right)
.\label{A3-7}%
\end{equation}
Of course, solution for antisymmetric and traceless field is antisymmetric
(RHS of (\ref{A3-7}) is manifestly antisymmetric) and traceless (contracting
(\ref{A3-7}) with $\delta_{m}^{z}$ or $\delta_{k}^{z}$).

At this stage, we can check the $D=3$ limit. The solution for $\hat{\Sigma
}_{z}^{\ \left(  mk\right)  }$ (\ref{A3-7}) was obtained for all dimensions
$D>2$ and it has to vanish when $D=3$. It is not difficult to check, taking
$\hat{\Sigma}_{z}^{\ \left(  mk\right)  }$ from (\ref{A3-7}) and using the
exact expressions of $\hat{D}_{\ \left(  bz\right)  }^{m}$ (\ref{A3-4}) and
$\hat{D}_{\ pq}^{m}$ (\ref{A3-1}), that such a limit is preserved%

\begin{equation}
\lim_{D=3}\hat{\Sigma}_{z}^{\ \left(  mk\right)  }\left(  \pi\right)
=\lim_{D=3}\hat{\Sigma}_{z}^{\ \left(  mk\right)  }\left(  e_{,s}\right)  =0.
\label{A3-8}%
\end{equation}

Actually, to demonstrate (\ref{A3-8}), it is not necessary to substitute the
explicit form of $\hat{D}_{\ \left(  bz\right)  }^{m}$. When $D=3$, there are
only two independent components of $\hat{\Sigma}_{z}^{\ \left(  mk\right)  }$
and they are $\hat{\Sigma}_{1}^{\left(  12\right)  }=\hat{\Sigma}_{2}^{\left(
12\right)  }=0$ because $\hat{\Sigma}_{z}^{\ \left(  mk\right)  }$ is
antisymmetric and traceless with only spatial indices.

Substitution of (\ref{A3-7}) back into (\ref{eqnD61}) or (\ref{eqnD75}) gives
the reduced Lagrangian or Hamiltonian, respectively.

\end{document}